\documentclass[prd,english,preprintnumbers,amsmath,amssymb,nofootinbib,twocolumn,superscriptaddress,aps,10pt]{revtex4-1}

\pdfoutput=1

\usepackage[utf8]{inputenc}
\usepackage{graphicx}
\usepackage{bbm}
\usepackage{amssymb}
\usepackage{amsmath}
\usepackage{tabularx}
\usepackage{slashed}

\usepackage{xspace}
\usepackage[export]{adjustbox}  

\usepackage{color}

\usepackage{dsfont}
\usepackage{babel}

\usepackage{hyperref}

\def\0#1#2{\frac{#1}{#2}}

\def\s0#1#2{\mbox{\small{$ \frac{#1}{#2} $}}}

\def\CC{{\mathcal C}}

\newcommand{\I}{\mathrm{i}}
\newcommand{\be}{\begin{eqnarray}}
\newcommand{\ee}{\end{eqnarray}}
\newcommand{\del}{\partial}

\newcommand{\nn}{\nonumber }

\newcommand{\beq}{\begin{equation}}
\newcommand{\eeq}{\end{equation}}
\newcommand{\bea}{\begin{eqnarray}}
\newcommand{\eea}{\end{eqnarray}}

\newcommand{\Nc}{N_{\rm c}}
\newcommand{\Nf}{N_{\rm f}}

\def\CC{{\mathcal C}}

\newcommand{\gs}{g_{\mathrm s}}

\newcommand{\UAone}{U_\mathrm{A}(1)}

\newcommand{\SpPm}{{(S+P)_{-}}}

\newcommand{\SpPmAdj}{{(S+P)_{-}^\mathrm{adj}}}
\newcommand{\Csc}{{\mathrm{csc}}}

\newcommand{\sigmapi}{(\sigma \text{-} \pi)}
\newcommand{\VmAPar}{{(V-A)_{\parallel}}}
\newcommand{\VmAPer}{{(V-A)_{\perp}}}
\newcommand{\VpAPar}{{(V+A)_{\parallel}}}
\newcommand{\VpAPer}{{(V+A)_{\perp}}}

\newcommand{\VmAPerAdj}{{(V-A)_{\perp}^{\mathrm{adj}}}}
\newcommand{\VpAParAdj}{{(V+A)_{\parallel}^{\mathrm{adj}}}}

\newcommand{\Euclidean}{\textit{Euclidean}\xspace}

\newcommand{\Dirac}{\textit{Dirac}\xspace}

\newcommand{\Fierz}{\textit{Fierz}\xspace}
\newcommand{\HubbardS}{\textit{Hubbard}-\textit{Stratonovich}\xspace}

\newcommand{\Ward}{\textit{Ward}\xspace}
\newcommand{\Landau}{\textit{Landau}\xspace}

\newcommand{\Takahashi}{\textit{Takahashi}\xspace}

\makeatother
\begin{document}

\title{Fierz-complete NJL model study III:\\ Emergence from quark-gluon dynamics}

\author{Jens Braun}
\affiliation{Institut f\"ur Kernphysik (Theoriezentrum), Technische Universit\"at Darmstadt, 
D-64289 Darmstadt, Germany}
\affiliation{ExtreMe Matter Institute EMMI, GSI, Planckstra{\ss}e 1, D-64291 Darmstadt, Germany}
\author{Marc Leonhardt} 
\affiliation{Institut f\"ur Kernphysik (Theoriezentrum), Technische Universit\"at Darmstadt, 
D-64289 Darmstadt, Germany}
\author{Martin Pospiech}
\affiliation{Institut f\"ur Kernphysik (Theoriezentrum), Technische Universit\"at Darmstadt, 
D-64289 Darmstadt, Germany}

\begin{abstract}
Our understanding of the dynamics and the phase structure 
of dense strong-interaction matter is to a large extent still built 
on the analysis of low-energy models, such as those of the 
{\it Nambu}--{\it Jona}-{\it Lasinio}-type. 
In this work, we analyze the emergence of the latter class of models at intermediate and 
low energy scales from fundamental quark-gluon interactions. 
To this end, we study the renormalization group flow of a {\it Fierz}-complete set of four-quark interactions 
and monitor their strength at finite temperature and quark chemical potential.
At small quark chemical potential, we find that the scalar-pseudoscalar interaction channel 
is dynamically rendered most dominant by the gauge degrees of freedom, indicating the formation of a chiral condensate.
Moreover, the inclusion of quark-gluon interactions leaves a significant imprint on the dynamics 
as measured by the curvature of the finite-temperature phase boundary which we find to be in accordance with lattice QCD results.
At large quark chemical potential, we then observe that 
the dominance pattern of the four-quark couplings is changed by the underlying quark-gluon dynamics, without any fine-tuning of the 
four-quark couplings. In this regime, the scalar-pseudoscalar interaction channel becomes subleading 
and the dominance pattern suggests the formation of
a chirally symmetric diquark condensate. In particular, our study confirms 
the importance of explicit~$U_\mathrm{A}(1)$ breaking for the formation 
of this type of condensate at high densities.
\end{abstract}

\maketitle
%%%%%%%%%%%%%%%%%%%%%%%%%%%%%%%%%%%%%%%%%%%%%%%%%%%%%%%%%%

% 
\section{Introduction}
Low-energy models of the theory of the strong
interaction (Quantumchromodynamics, QCD) are still considered very valuable 
for a variety of reasons. In particular in the high-density regime, which is 
at least difficult to access with lattice Monte Carlo techniques, 
the {\it Nambu--Jona-Lasinio} (NJL) model~\cite{Nambu:1961tp,Nambu:1961fr}
and its various variations and relatives (see, e.g., Refs.~\cite{Klevansky:1992qe,Hatsuda:1994pi,Berges:2000ew,Schaefer:2006sr,vonSmekal:2012vx,Drews:2016wpi} for 
reviews), 
such as quark-meson (QM) models, allow us 
to gain some insight into the plethora of symmetry breaking patterns that may potentially be 
realized in this regime, see Refs.~\cite{Buballa:2003qv,Fukushima:2011jc,Alford:2007xm,Anglani:2013gfu} for reviews. 
From a phenomenological point of view, this regime attracts significant interest from the 
astrophysics side. For example, 
studies of neutron stars require a knowledge of the equation of state of strong-interaction matter as input. 
However, the latter are currently
still plagued by (significant) uncertainties at least at high density, see, e.g., Ref.~\cite{Watts} 
for a recent review. In order to constrain the equation of state further~\cite{Leonhardt:2019fua}, we 
eventually need a better understanding of the symmetry breaking patterns of QCD guided by 
first-principle approaches. 

In two preceding works~\cite{Braun:2017srn,Braun:2018bik}, 
we have studied the relevance of \Fierz completeness of four-quark self-interactions in 
NJL-type models at finite temperature and quark chemical potential. 
Our beyond mean-field 
renormalization-group (RG) analysis of the ``hierarchy'' of the various interaction channels in terms of their relative strengths 
allowed us to gain insight into the symmetry breaking patterns and the 
structure of the ground state. Particularly at high density, 
we found the aspect of \Fierz completeness to be of great importance, 
leading to an increased phase transition temperature as compared to conventional NJL model studies. 
At least naively, this observation might have crucial implications for the properties of cold dense quark matter 
at low temperatures as an increase of the critical temperature may point to 
an increase of the size of the gap in the low-energy spectrum. 

The four-quark couplings appearing in the ansatz of NJL-type models are usually considered 
as fundamental parameters. In fact, owing to the non-renormalizability of NJL-type models in 
four space-time dimensions, both on the perturbative as well as on the non-perturbative level (see, e.g., Refs.~\cite{Braun:2011pp,Braun:2012zq}), 
the ultraviolet (UV) cutoff scale~$\Lambda$ becomes a parameter of the model, too. Against this background, the regularization scheme becomes 
also part of the definition of the model. 
The initial values of the four-quark couplings are then chosen such that a given set of low-energy observables is reproduced in the vacuum limit. 
In Ref.~\cite{Braun:2018bik}, guided by the findings of RG studies 
of QCD~\cite{Braun:2006zz,Mitter:2014wpa,Cyrol:2017ewj} and in order to relate to conventional NJL model studies, 
we have set all four-quark couplings to zero initially, except for the scalar-pseudoscalar coupling. 
As the only remaining parameter, we adjusted this coupling to fix low-energy observables
in the vacuum limit.
However, this scale fixing procedure underlying many low-energy model studies can be potentially problematic. 
The distinct role of the scalar-pseudoscalar interaction channel at the initial UV cutoff scale can be indeed 
questioned since a specific four-quark interaction channel is reducible by means of \Fierz transformations. 
Yet adopting more complex initial conditions by also taking into account four-quark couplings other than the 
scalar-pseudoscalar interaction channel 
may suffer from the fact that the parameters cannot be uniquely determined by a given set of low-energy observables. 
Indeed, the values of the low-energy observables may in general be reproduced by various different 
parameter sets or certain parameters might be even left undetermined at all. The remaining ambiguity has been 
found to affect the phase structure significantly, see, e.g., Refs.~\cite{Hatsuda:1985eb,Buballa:2003qv,Schaefer:2008hk,Zhang:2009mk}. 
Moreover, boundary conditions which are defined in the vacuum limit are possibly inappropriate for 
computations at finite external control parameters, such as temperature~$T$ and/or quark chemical potential~$\mu$. 
Considering NJL-type models to be rooted in QCD, the RG evolution of gluon-induced four-quark interactions in fact 
suggests a dependence of these model parameters on external control parameters~\cite{Springer:2016cji}.
In particular at finite quark chemical potential, as also discussed in detail in Refs.~\cite{Braun:2017srn,Braun:2018bik} 
(see, e.g., Refs.~\cite{Buballa:2003qv,Bailin:1983bm,Alford:2007xm,Anglani:2013gfu} for reviews), 
effective degrees of freedom associated with four-quark interaction channels other than the scalar-pseudoscalar interaction channel 
are expected to become important or even dominant. 
In such a situation, a choice for the initial conditions with a specifically tuned scalar-pseudoscalar coupling 
is therefore unfortunate as it may potentially bias 
the outcome in terms of symmetry breaking patterns along the finite-temperature phase boundary.

Thus far, we have not yet discussed the role of the UV cutoff scale.
In the context of NJL-type models, we have to deal with the existence of a finite UV extent, 
i.e., the cutoff scale~$\Lambda$ is limited by a validity bound which in turn limits the model's range of applicability in 
terms of external parameters. The origin of this bound is actually twofold:
First, NJL-type models eventually become unstable in the UV and develop a \Landau 
pole at a certain scale. Second, these models have a phenomenological UV extent beyond which  
the description of the physics in terms of the models' effective fields becomes invalid and 
the knowledge of the underlying fundamental dynamics, i.e., quark-gluon dynamics, is ultimately required. 
As a consequence, a choice of the UV cutoff scale within the {\it a priori} unknown validity bound
either limits the applicable range of external parameters or, for external parameters outside of this range, implies that the initial
effective action is already a complicated object itself, see Ref.~\cite{Braun:2018svj} for a detailed discussion 
of these issues. 
Considering NJL-type models to be embedded in QCD, 
a possibility to resolve this problem is the 
determination of the model parameters by employing RG studies of the fundamental theory, 
see, e.g., Refs.~\cite{Braun:2003ii,Herbst:2013ufa,Springer:2016cji}.
In the past 20 years, tremendous progress has been made within the functional RG framework 
in the development of a ``top-down'' approach to QCD, see, e.g., Refs.~\cite{Meggiolaro:2000kp,Gies:2002af,Gies:2002hq,Pawlowski:2003hq,%
Fischer:2004uk,Gies:2005as,Braun:2005uj,Braun:2006jd,%
Braun:2007bx,Braun:2008pi,Marhauser:2008fz,%
Braun:2009gm,Braun:2010cy,Kondo:2010ts,Fister:2013bh,Braun:2014ata,Mitter:2014wpa,%
Rennecke:2015eba,Cyrol:2016tym,Cyrol:2017ewj,Cyrol:2017qkl,Corell:2018yil,Fu:2019hdw}. 
The only input in such an approach 
 is given by the fundamental parameters of QCD, i.e., the current quark masses and the value of the strong coupling 
 in the perturbative high-momentum regime. These functional RG studies are basically free of additional 
 model parameters. In this context, even very good quantitative agreement of results from lattice QCD and functional RG studies
 has been demonstrated at zero and finite temperature for QCD with different flavor numbers, 
 see, e.g., Refs.~\cite{Gies:2005as,Braun:2006jd,Braun:2007bx,Cyrol:2017ewj,Fu:2019hdw}.
 
 Whereas the aforementioned RG studies aiming at quantitative precision constitute 
 essential advances towards predictive first-principle investigations of the QCD phase diagram, eventually even at high densities, 
we aim to take another important step toward such a top-down first-principle approach to analyze the phase structure of QCD at high densities 
with functional methods. 
With our analysis of the \Fierz-complete NJL model with two quark flavors in Ref.~\cite{Braun:2018bik},
we have gained valuable insight into the quark dynamics. 
In the present work,  
we now incorporate gluodynamics by extending our \Fierz-complete ansatz with dynamical gauge degrees of freedom, following 
earlier functional RG studies~\cite{Braun:2005uj,Braun:2006jd,Braun:2008pi,Braun:2014ata}. 
In full QCD, the values of the four-quark couplings are no longer fundamental parameters 
since these self-interactions are fluctuation-induced by the dynamics of the gauge fields. 
Taking this aspect into account, the aforementioned issue associated with the determination of model parameters  
-- such as ambiguities related to the possibility to \Fierz-transform given initial conditions and the potential existence of more 
than one parameter set reproducing equally well a given set of low-energy observables, or the dependence of the initial 
conditions on external control parameters -- can in principle be resolved. More specifically, 
including gauge dynamics and thus resolving the fundamental microscopic degrees of freedom allows the initialization 
of the RG flow at a large scale~$\Lambda$ associated with the perturbative regime,  
which effectively corresponds to starting in the vacuum as we have $T / \Lambda \ll 1$ and $\mu / \Lambda \ll 1$. 
In this way, the finite UV extent as implied by the validity bound of NJL-type models is 
surmounted and the limit on the range of applicability in terms of external parameters is lifted. 

Working in the chiral limit, the strong coupling $\gs$ is the only parameter which is set at a the initial UV scale~$\Lambda$.  
By integrating out fluctuations from this scale~$\Lambda$ on, 
the quark-gluon vertex gives rise to $1$PI box diagrams with two-gluon exchange which dynamically generate the four-quark interaction channels. 
Depending on the strength of the strong coupling and the external parameters, the quark sector 
may then be driven to criticality, signaling the onset of spontaneous symmetry breaking, e.g., chiral symmetry breaking or diquark condensation 
of a specific type. 
Following the approach developed in our two preceding works~\cite{Braun:2017srn,Braun:2018bik}, 
we shall consider the RG flow of the four-quark couplings in the pointlike limit to study the QCD phase structure at finite temperature 
and quark chemical potential. 
In particular, we analyze the ``hierarchy'' of the four-quark couplings in terms of their strength 
which shall prove very valuable in order to gain insight into the symmetry properties of the QCD ground state in the low-energy limit. 

This work is organized as follows: In Sec.~\ref{sec:genasp}, we discuss general aspects of the formalism and 
concepts underlying our study. We begin 
with a discussion of the relation of the quark-gluon vertex and four-quark interactions in Subsec.~\ref{sec:app}. 
In Subsec.~\ref{subsec:pt}, we then briefly review the relation of the RG flow of four-fermion couplings to the onset of phase transitions, including 
a discussion of the general structure of the RG flow equations for the four-quark couplings. 
The scale fixing procedure underlying our present work is discussed in Subsec.~\ref{sec:sfproc}.
The QCD phase structure and symmetry breaking patterns at finite temperature and density 
are then analyzed in Sec.~\ref{sec:fpps}. 
There, we also compare the results for the phase boundary to the one obtained from our previous 
\Fierz-complete NJL model study~\cite{Braun:2018bik}. Moreover,  
we discuss the effect of explicit $\UAone$ symmetry breaking and 
comment on the curvature of the finite-temperature phase boundary at small chemical potential resulting from various different approaches.
Our conclusions can be found in Sec.~\ref{sec:conc}.

\section{General Aspects of the Formalism}\label{sec:genasp}
\subsection{Quark-gluon vertex and four-quark interactions}\label{sec:app}
In the present work, we employ the functional
RG approach~\cite{Wetterich:1992yh} to 
study the RG flow of QCD starting from  
the \Euclidean QCD action (see
Refs.~\cite{Pawlowski:2005xe,Gies:2006wv,Braun:2011pp} for reviews):
\be
S = \int {\rm d}^4x\left\{ 
\frac{1}{4}F_{\mu\nu}^{a}F_{\mu\nu}^{a}
+ \bar{\psi}\left(
{\rm i}\slashed{\partial} + \bar{g}_{\rm s}\slashed{A} +{\rm i}\gamma_0\mu \right)\psi
\right\}\,,
\label{eq:qcd}
\ee
where~$\bar{g}_{\rm s}$ is the bare gauge coupling and~$\mu$ is the quark chemical
potential. The gluon fields~$A_{\mu}^a$ enter the definition of the
field-strength tensor~$F_{\mu\nu}^a$ in the usual way. 
We emphasize that we exclusively consider the case of quarks coming in~$\Nc=3$ colors and $\Nf=2$ flavors. 

In the RG flow, the quark-gluon vertex generates quark
self-interactions already at the one-loop level via two-gluon exchange. This
gives rise to contributions to the quantum effective action, e.g., of the following form:
\be
\delta\Gamma = \frac{1}{2}\int {\rm d}^4x\,\sum_{j \in\, \mathcal{B}}\ Z_j{\bar\lambda_{j}}\,\mathcal{L}_{j} \,,
\label{eq:deltag}
\ee
where the elements $\mathcal{L}_{j}$ form a
ten-component {\it Fierz}-complete basis~${\mathcal B}$ of pointlike four-quark interactions to be specified below. 
The various terms are associated with corresponding bare couplings~$\bar{\lambda}_i$ and vertex renormalization factors~$Z_j$. 
By construction, the couplings are not parameters of our calculation but solely generated by
quark-gluon dynamics. This is an important
difference to, e.g., NJL-type model studies where such
couplings represent input parameters.

In the following, we focus on the RG flow of pointlike projected four-quark
correlation functions~$\Gamma^{(4)}$ which eventually corresponds to a calculation of the 
effective action at leading order of the derivative expansion. To be specific, we define the four-quark
couplings associated with the vertex of the form~\eqref{eq:deltag} as follows
\be
\label{eq:def4qc}
&&Z_j{\bar\lambda_{j}}\,\mathcal{L}_{j} \\
&&  =\! \lim_{p_i\to 0}
\bar\psi_{\alpha}(p_1) \bar\psi_{\beta}(p_2)
\Gamma^{(4), \alpha\beta\gamma\delta}_{\mathcal{L}_{j}}\! (p_1,p_2,p_3,p_4) \psi_{\gamma}(p_3) \psi_{\delta}(p_4).\nn
\ee
Here, $\alpha, \beta, \gamma, \delta$ denote collective indices for color,
flavor, and \Dirac structures determined by a specific element~$\mathcal{L}_{j}$ of our \Fierz-complete basis.
We add that this zero-momentum projection does not
represent a {\it Silver}-{\it Blaze}-symmetric point~\cite{Khan:2015puu,Fu:2016tey,Braun:2017srn}. However, 
it matches the standard definition of four-quark couplings in conventional low-energy
models (see Refs.~\cite{Klevansky:1992qe,Hatsuda:1994pi,Buballa:2003qv,Fukushima:2010bq,Fukushima:2011jc} for reviews) and BCS-type
models (see Refs.~\cite{Bailin:1983bm,Altland:2006si,Alford:2007xm,Anglani:2013gfu} for reviews). 

Let us now specify the elements $\mathcal{L}_{j}$ of our {\it Fierz}-complete basis~$\mathcal B$ of pointlike 
four-quark interactions. 
Since {\it Poincar\'{e}} invariance is explicitly broken in our calculations at finite temperature 
and quark chemical potential, we are only left with rotational invariance among the spatial components of the various possible channels. 
Moreover, a finite quark chemical potential also explicitly breaks the charge conjugation symmetry. Therefore, 
with respect to the fundamental symmetries associated with charge conjugation, time reversal,
and parity, only invariance under parity transformations and time reversal transformations
remain intact. Assuming finally invariance of the channels 
under~$SU(\Nc)\otimes SU_\text{L}(2)\otimes SU_\text{R}(2)\otimes U_\text{V}(1)$, we end up  
with a \Fierz-complete basis composed of ten elements~\cite{Braun:2018bik}.
Guided by QCD low-energy phenomenology,  
we choose four of the ten channels such that they are invariant {under~$SU(\Nc)\otimes SU_\text{L}(2)\otimes SU_\text{R}(2)\otimes U_\text{V}(1)$ 
transformations}
but break the $U_\text{A}(1)$ symmetry explicitly:
\be
\mathcal{L}_{\text{($\sigma $-$\pi $)}}&=&\left(\bar\psi \psi\right)^2\!-\! \left(\bar\psi \gamma_5 \tau_i \psi\right)^2\,,\label{eq:SNJL} 
\\
\mathcal{L}_{(S+P)_{-}}&=&\left(\bar\psi \psi\right)^2\!-\!\left(\bar\psi \gamma_5 \tau_i \psi\right)^2 \nn\\
&&\hspace{1cm} \!+\!\left(\bar\psi \gamma_5 \psi\right)^2\!-\!\left(\bar\psi \tau_i \psi\right)^2\,,\\
\mathcal{L}_\Csc &=& 4 \left( \I \bar \psi \gamma_5 \tau_2\, T^{A} \psi^C \right) \left( \I \bar \psi^C \gamma_5 \tau_2\, T^{A} \psi \right)\,,
\label{eq:cscdef}\\
\mathcal{L}_{(S+P)_{-}^\mathrm{adj}} &=&\left(\bar\psi T^a\psi\right)^2\!-\!\left(\bar\psi \gamma_5 \tau_i T^a\psi\right)^2 \nn\\
&&\hspace{1cm}\!+\!\left(\bar\psi \gamma_5 T^a\psi\right)^2\!-\!\left(\bar\psi \tau_i T^a\psi\right)^2\,,
\label{eq:detcp}
\ee
where, e.g., $\left(\bar\psi \gamma_5 \tau_i \psi\right)^2\equiv \left(\bar\psi \gamma_5 \tau_i \psi\right)\left(\bar\psi \gamma_5 \tau_i \psi\right)$ 
and the $T^a$'s denote the generators of $SU(\Nc)$. Moreover, we 
introduced charge conjugated fields~$\psi^C = \CC \bar\psi ^T$ and $\bar \psi^C = \psi^T \CC $ 
with $\mathcal{C}=\I \gamma_2 \gamma_0$ being related to the charge conjugation operator.
The remaining six channels can then be chosen to be even invariant 
under $SU(\Nc)\otimes SU_\text{L}(2)\otimes SU_\text{R}(2)\otimes U_\text{V}(1) \otimes U_\text{A}(1)$ transformations:
\be
\mathcal{L}_\VpAPar&=&\left(\bar\psi\gamma_0\psi\right)^2+\left(\bar\psi\I\gamma_0\gamma_5\psi\right)^2\,,\label{eq:firstchan}
\\
\mathcal{L}_\VpAPer&=&\left(\bar\psi\gamma_i\psi\right)^2+\left(\bar\psi\I\gamma_i\gamma_5\psi\right)^2\,,\\
\mathcal{L}_\VmAPar&=&\left(\bar\psi\gamma_0\psi\right)^2-\left(\bar\psi\I\gamma_0\gamma_5\psi\right)^2\,,\\
\mathcal{L}_\VmAPer&=&\left(\bar\psi\gamma_i\psi\right)^2-\left(\bar\psi\I\gamma_i\gamma_5\psi\right)^2\,,\\
\mathcal{L}_\VpAParAdj&=&\left(\bar\psi\gamma_0 T^a\psi\right)^2+\left(\bar\psi\I\gamma_0\gamma_5 T^a\psi\right)^2\,,
\label{Eq:VpAParAdj}\\
\mathcal{L}_\VmAPerAdj&=&\left(\bar\psi\gamma_iT^a\psi\right)^2-\left(\bar\psi\I\gamma_i\gamma_5 T^a\psi\right)^2\,.
\ee
Of course, this basis is not unique. In principle, we can combine elements of the 
basis to perform a basis transformation. However, as indicated above, our present choice is motivated 
by the structure of the four-quark channels conventionally employed in QCD low-energy models. In fact, 
the scalar-pseudoscalar channel associated with pion dynamics and chiral symmetry breaking 
is given by the channel~$\mathcal{L}_{\text{($\sigma $-$\pi $)}}$. 
The channel associated with the element $\mathcal{L}_{(S+P)_{-}}$ can be rewritten as 
(up to a numerical constant)
\be
\sim \det\left(\bar{\psi}(1+\gamma_5)\psi\right) + \det\left(\bar{\psi}(1-\gamma_5)\psi\right)\,,
\label{eq:det}
\ee
where the determinant is taken in flavor space. This channel is associated 
with the presence of topologically non-trivial gauge configurations violating the~$U_{\text{A}}(1)$ symmetry.  
Indeed, such configurations can be 
recast into a four-quark interaction channel of the form~\eqref{eq:det} in the case of 
two-flavor QCD~\cite{tHooft:1976snw,tHooft:1976rip,Shifman:1979uw,Shuryak:1981ff,Schafer:1996wv,Pawlowski:1996ch,Kunihiro:2010rj}.
The channel~\eqref{eq:detcp} may be 
viewed as a version of the channel~$\mathcal{L}_{(S+P)_{-}}$ with a non-trivial
color structure. 
We add that, for the phenomenologically more relevant three-flavor case, 
the explicit breaking of the $\UAone$ symmetry has been suggested already before the two-flavor case in order 
to explain the mass splitting of the $\eta$ and $\eta^{\prime}$ in the mesonic mass spectrum~\cite{Kobayashi:1970ji,Kobayashi:1971qz}.

Finally, we have included a channel 
which allows us to ``measure" the status of the formation of a diquark condensate of the type 
\be
\delta^{a} \sim\langle \I \bar \psi^C \gamma_5 \epsilon_{f}\varepsilon_{c}^a\psi \rangle\,,
\ee
which carries a net baryon and net color charge.\footnote{Here,~$\epsilon_{f}\equiv \epsilon_f^{(\alpha,\beta)}$ 
and $\varepsilon_{c}^{a}\equiv\varepsilon_c^{a(m,n)}$ are 
antisymmetric tensors in flavor and color space, respectively.}
Note that the channel~$\mathcal{L}_\Csc$ 
is invariant under~$SU(\Nc)\otimes SU_\text{L}(2)\otimes SU_\text{R}(2)\otimes U_\text{V}(1)$ transformations and the
corresponding condensate leaves the chiral symmetry intact. However, the formation of such a diquark 
condensate comes along with the breakdown 
of the~$U_\text{V}(1)$ symmetry, as expected for a BCS-type condensate.\footnote{With respect to the diquark channel, we add that 
our conventions in Eq.~\eqref{eq:cscdef} are such 
that we only sum over the antisymmetric~($A$) generators {of the $SU(\Nc)$ color} group. Normalization factors of this channel are 
chosen according to the standard literature~\cite{Buballa:2003qv}.}

With respect to our discussion of the effect of $U_{\text{A}}(1)$ symmetry breaking, 
we add that we can use our {\it Fierz}-complete set of pointlike four-quark interactions to
monitor the strength of $U_{\text{A}}(1)$ symmetry breaking.
Indeed, requiring that the effective action~$\Gamma$ is invariant under~$U_{\text{A}}(1)$ transformations, we find the following 
two sum rules which are satisfied simultaneously if the $U_{\text{A}}(1)$ symmetry is intact:
\be
\!\!\!\!\!\!\!\!\!\!\mathcal{S}_{U_{\rm A}(1)}^{(1)}&=& \bar{\lambda}_\Csc+ \bar{\lambda}_\SpPmAdj =0\,,\label{eq:SUA1}\\
\!\!\!\!\!\!\!\!\!\!\mathcal{S}_{U_{\rm A}(1)}^{(2)}&=& \bar{\lambda}_\SpPm \!-\! \frac{\Nc \!-\! 1}{2\Nc}\,\bar{\lambda}_\Csc \!+\!\frac{1}{2}\bar{\lambda}_{(\sigma \text{-} \pi)} =0\,,
\label{eq:SUA2}
\ee
see Ref.~\cite{Braun:2018bik} for a more detailed discussion. 
We shall come back to the issue of~$U_{\text{A}}(1)$ symmetry breaking in Subsec.~\ref{sec:GaugeUA1}.  

To summarize, from here on, we shall consider the RG flow of pointlike four-quark interactions as generated by 
the quark-gluon vertex. The latter implies that the initial conditions of the four-quark interactions are set to zero 
at the initial RG scale~$\Lambda$, i.e., they do not represent parameters of our study as it is the case for 
QCD low-energy models. We emphasize again that our set of four-quark interactions  
is \Fierz-complete, i.e., 
any other pointlike four-quark interaction invariant under the symmetries specified above 
 is reducible by means of {\it Fierz} transformations. 
 
 Since we consider the RG flow of the four-quark interactions at leading 
 order of the derivative expansion (i.e., in the pointlike limit as indicated above),
 the quark wave-function renormalization factors 
 do not receive contributions directly from the four-quark interactions, 
 see Ref.~\cite{Braun:2011pp} for a detailed discussion. Contributions 
 to the quark wave-function renormalizations resulting
 from the coupling of the quarks to the gluons 
 have been found to be small at this order in earlier studies~\cite{Gies:2002hq,Gies:2005as,Braun:2008pi}, 
 at least for RG scales relevant for the present work. Therefore, 
 we do not take into account the 
 running of the quark wave-function renormalization factors and defer it to future work.  
Finally, we note that quark self-interactions of higher order are also generated 
dynamically in the RG flow. However, at leading order of the derivative expansion, they do
not contribute to the RG flow of the four-quark self-interactions  
and are thus not included in our present study. In any case, 
since the RG flow equations of the four-quark couplings depend on the strong coupling, we 
also need to take into account the running of the strong coupling which we shall discuss 
below.

\subsection{Phase transitions and four-quark interactions}\label{subsec:pt}
The four-quark couplings depend on the chemical potential, the temperature, and the RG
scale~$k$. Although the scale dependence implies that part of the
information on the momentum dependence is still taken into account in our RG analysis
in an effective manner~\cite{Braun:2014wja}, the pointlike approximation underlying our 
present work ignores
relevant information of four-quark correlation functions. To be more specific,
bound-state information is encoded in the momentum structure of the quark correlation functions. 
Therefore, our present approximation only allows us
to study the symmetric high-energy regime~\cite{Braun:2011pp} whereas the 
low-energy regime is only accessible in the absence of (spontaneous) symmetry breaking which is 
associated with bound-state formation, as it is the case at, e.g., high temperature. For our
purposes, this is nevertheless sufficient as it
enables us to study the approach towards the symmetry-broken low-energy regime.
Indeed, symmetry breaking is ultimately triggered by a specific four-quark
channel approaching criticality as indicated by a divergence of the
corresponding coupling. Such a seeming \Landau-pole-type behavior of four-quark
couplings can be traced back to the formation of condensates since 
the pointlike four-quark couplings can be shown to be proportional to 
the inverse mass parameters of a generalized {\it Ginzburg}-{\it Landau}
effective potential for the order parameters in a (partially) bosonized
formulation~\cite{Ellwanger:1994wy,Gies:2001nw,Braun:2011pp}.
Thus, if the size of all
four-quark couplings is found to be bounded in the RG flow, the system remains in
the symmetric regime~\cite{Gies:2005as,Braun:2005uj,
Braun:2006jd,Braun:2011pp,Braun:2014wja}. The observation
of a divergent four-quark coupling for a given temperature and quark chemical potential may therefore be
considered as an indicator that the order-parameter potential develops a
non-zero ground-state expectation value in the direction associated with a specific
four-quark channel. 
The critical temperature~$T_\mathrm{cr}(\mu)$ at a given value of the quark chemical potential above which 
no spontaneous symmetry breaking occurs is then defined as the smallest temperature for which the four-quark couplings still remain 
finite in the infrared (IR) limit associated with $k\to 0$~\cite{Braun:2005uj, Braun:2006jd,Braun:2011pp,Braun:2017srn,Braun:2018bik}. 
However, we emphasize that our present approach is only able to detect phase transitions of second order as the definition of 
the critical temperature is associated with a change from positive to negative curvature of the order parameter potential 
at the origin. In case of a first-order phase transition, a non-trivial minimum of the potential is 
formed but the curvature at the origin remains positive. Consequently, 
our criterion for the detection of a phase transition does not allow to detect first-order transitions. Still, it 
allows us to detect the line of metastability~\cite{Braun:2017srn}, see 
for the first NJL model analysis of this aspect~\cite{Asakawa:1989bq}.
In any case, the nontrivial assumption entering our analysis 
of the QCD phase structure in the present work 
is that it is possible to relate the dominance pattern of the
four-quark couplings to the symmetry-breaking pattern in terms of condensates,
see Refs.~\cite{Braun:2014wja,Braun:2017srn,Braun:2018bik,Roscher:2019omd} for a detailed
discussion of this aspect. 

It is reasonable to expect that
the symmetry-breaking patterns associated with the various four-quark channels
change when we vary, e.g., the quark chemical potential. 
For example, the diquark channel may become more relevant than the scalar-pseudoscalar
channel at high density. The most dominant channel can be
identified by requiring that the modulus of the coupling of this channel is
greater than the ones of all the other four-quark couplings. For such an analysis 
to be meaningful, it is therefore ultimately required to consider a \Fierz-complete 
set of four-quark couplings. 
\begin{figure}[t]
\centering
\includegraphics[width=\linewidth]{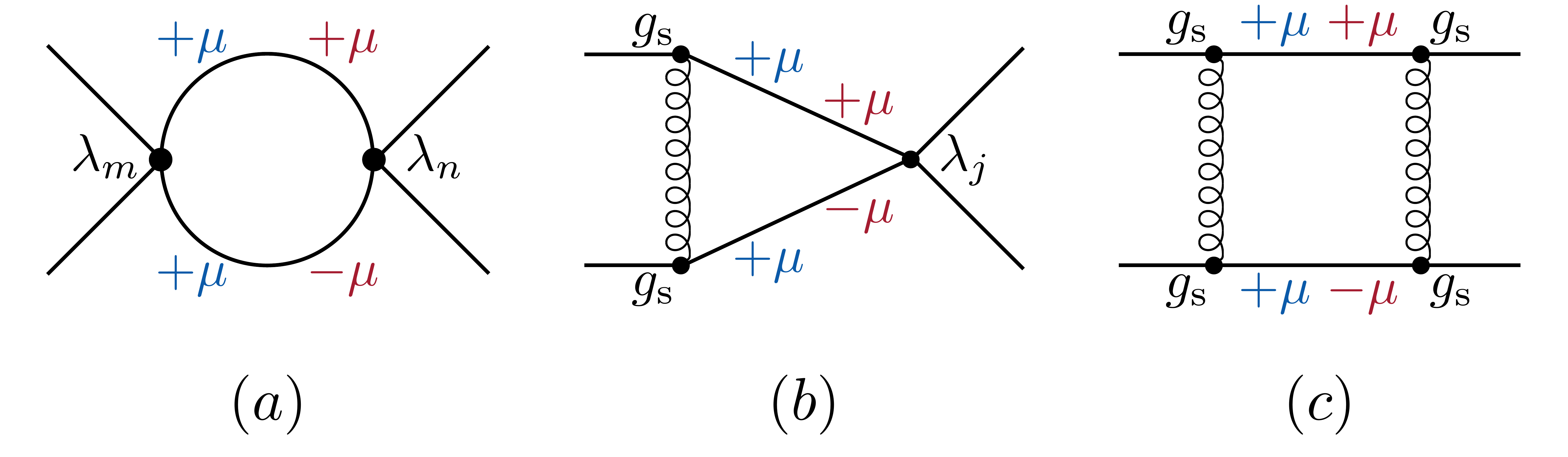} 
\caption{$1$PI diagrams contributing to the RG flow~\eqref{eq:GaugeStructureBetaFunctions} of the four-quark 
couplings~$\lambda_i$. Note that there are two classes of each diagram which are associated with the signs of the quark chemical potential appearing 
in the quark propagators: one class represents the case of equal signs as depicted by the blue labels and 
the other one represents the case of opposite signs as depicted by the red labels.
}
\label{fig:GaugeDiagrams}
\end{figure}

For the derivation of the RG flow equations of the four-quark couplings, we employ 
the {\it Wetterich} equation~\cite{Wetterich:1992yh} which is an RG equation for the quantum effective action~$\Gamma$.
Within this framework, the effective action~$\Gamma$ depends on the RG scale~$k$ which is 
related to the so-called RG ``time" $t=\ln (k/\Lambda)$. Note that the scale~$k$ defines an infrared (IR) cutoff scale 
and~$\Lambda$ may be chosen to be the scale at which
we fix the initial conditions of the RG flow of the couplings.
The general structure of the RG flow equations 
for the dimensionless renormalized couplings $\lambda_i = Z_i k^2 \bar \lambda_i$ is given by
\be
\del_t \lambda_ i &=& 2 \lambda_i - \mathcal{A}^{(i)}_{mn}(\tau, \tilde \mu_\tau) \lambda_m \lambda_n \nn \\
&& \quad - \mathcal{B}^{(i)}_j(\tau, \tilde \mu _\tau) \lambda_j \gs^2 - \mathcal{C}^{(i)}(\tau, \tilde \mu _\tau) \gs ^4 \, ,
\label{eq:GaugeStructureBetaFunctions}
\ee
with $\tau = T / k$ and $\tilde \mu _\tau = \mu /(2 \pi T)$. The temperature- and chemical-potential-dependent 
coefficients $\mathcal A^{(i)} _{mn}$, $\mathcal B^{(i)} _j$ and $\mathcal C^{(i)}$ are auxiliary functions containing 
sums of the $1$PI diagrams depicted in Fig.~\ref{fig:GaugeDiagrams}. 
In addition, these functions also contain the information on the chosen gauge. For convenience, we shall restrict ourselves to 
{\it Feynman} gauge in the following.
In Eq.~\eqref{eq:GaugeStructureBetaFunctions}, terms bilinear in the four-quark couplings with the coefficients $\mathcal{A}^{(i)}_{mn}$ 
are associated with the purely fermionic diagrams~(a) in Fig.~\ref{fig:GaugeDiagrams}. 
Terms proportional to $\lambda_j \gs ^2$ are generated by the triangle diagrams~(b) 
depicted in Fig.~\ref{fig:GaugeDiagrams}. Finally, 
terms proportional to $\gs ^4$ are associated with the box diagrams (c) shown in Fig.~\ref{fig:GaugeDiagrams}. 
We add that we have dropped an explicit dependence of these loop diagrams on the
anomalous dimensions of the quark and gluon fields as these additional contributions 
have been found to be subleading in the symmetric
regime~\cite{Gies:2002hq,Gies:2003dp,Gies:2005as,Braun:2005uj, Braun:2006jd}.

For the derivation of the set of RG flow equations~\eqref{eq:GaugeStructureBetaFunctions}, we have made
extensive use of existing software packages~\cite{Huber:2011qr,Cyrol:2016zqb}. 
Because of the size of the resulting system of equations for the \Fierz-complete set of four-quark couplings,  
we dealt with the flow equations only numerically and 
therefore refrain from listing these equations explicitly here.
An explicit representation of the flow equations for the purely fermionic part as parameterized by the
matrices~$\mathcal{A}^{(i)}_{mn}$ can be found in our preceding work~\cite{Braun:2018bik}, 
including a discussion of the regularization scheme also underlying our present work. 
For the same regularization scheme,  
an explicit representation of a \Fierz-complete set of flow equations for the four-quark interactions in the vacuum 
limit in 
case of an $SU(\Nc)\otimes U_\text{L}(2)\otimes U_\text{R}(2)$ symmetry can be found in 
Ref.~\cite{Gies:2005as}, where the contributions 
proportional to~$\lambda_j g_{\rm s}^2$ and~$ g_{\rm s}^4$ have been taken into account as well.

In our present approximation, the RG flow of the gauge sector enters the flow
equations of the four-quark couplings only via the running of the strong
coupling. In our numerical analysis in Sec.~\ref{sec:fpps}, we employ the 
running coupling computed non-perturbatively in Refs.~\cite{Braun:2005uj,Braun:2006jd} where the same
regularization scheme has been used as in the present work. 
Let us be more explicit at this point. In Refs.~\cite{Braun:2005uj,Braun:2006jd}, the running of 
the gauge coupling has been computed at zero and finite temperature, but not at finite quark chemical 
potential. An extension of these non-perturbative studies to finite quark chemical potential 
is beyond the scope of the present work and therefore deferred to future work. To estimate 
the effect of a quark chemical potential on the running of the strong coupling and thereby on 
the flow of the four-quark couplings, we employ the running strong coupling from Refs.~\cite{Braun:2005uj,Braun:2006jd} 
as obtained for (pure) {\it Yang}-{\it Mills} (YM) theory ($\Nf=0$) and for QCD with two massless 
quark flavors. The reason behind is that, at the one-loop level, the quark contribution 
to the $\beta$ function of the strong coupling at finite quark chemical potential and zero temperature has been found to be identical 
to the one in the vacuum limit  
for RG scales~$k \geq \mu$ and to vanish identically for~$k<\mu$, 
see, e.g., Refs.~\cite{Braun:2009gm,SpringerThesis,Fu:2019hdw}. This implies that, at least at the one-loop level at zero temperature, 
the RG flow of the strong coupling is identical to the one in the vacuum limit for~$k \geq\mu$ but 
identical to the one in YM theory for~$k<\mu$. Loosely speaking, the YM coupling and the
QCD coupling in the vacuum limit may therefore be viewed as two extremes of 
the zero-temperature QCD coupling at finite quark chemical 
potential. The stepwise change of the $\beta$ function of the strong coupling 
at~$k=\mu$ is then smeared out at finite temperature.

In addition to corrections to the strong coupling originating from the presence of 
a finite quark chemical potential, one may be worried about possible 
corrections to the strong coupling arising from the presence of quark self-interactions.
However, provided the
flow of the four-quark couplings is governed by the presence of fixed
points~\cite{Gies:2003dp}, as it is the case in the symmetric
regime~\cite{Gies:2005as,Braun:2005uj, Braun:2006jd}, it follows 
from the
analysis of (modified) \Ward-\Takahashi identities that
the back-reaction of the four-quark couplings on the strong coupling is negligible. 
For our present analysis, the use of the strong coupling from Refs.~\cite{Braun:2005uj,Braun:2006jd} 
in our set of flow equations for the four-quark couplings is therefore not only a convenient 
but also a consistent approximation. 

We close this discussion by adding that the mechanism underlying 
the dynamical generation of the four-quark couplings and 
the role of the running gauge coupling for spontaneous symmetry breaking can be already understood in 
simple terms by analyzing analytically the fixed-point structure of the RG flow equations for the four-quark couplings. 
Here, we refrain from repeating this general line of arguments and only refer 
the reader to the discussions given in 
Refs.~\cite{Gies:2005as,Braun:2005uj,Braun:2006jd,Braun:2009ns,Braun:2010qs,Kusafuka:2011fd}, see Ref.~\cite{Braun:2011pp} 
for an introduction. 

\subsection{Scale-fixing procedure}\label{sec:sfproc}
As already mentioned above, the initial values of the four-quark interaction channels are not considered 
as fundamental parameters in our present approach and are set to zero at the UV scale~$\Lambda$.
This corresponds to a $\UAone$-symmetric scenario according to the sum rules~\eqref{eq:SUA1} and~\eqref{eq:SUA2}. 
The case with explicit $\UAone$-symmetry breaking is discussed separately in Subsec.~\ref{sec:GaugeUA1}.
In any case, the four-quark couplings must be generated dynamically by quark-gluon interactions which removes a potential bias in the parameter choice 
as present in low-energy model studies. 
In fact, within our present study, we rather predict the values of the four-quark couplings from the underlying quark-gluon dynamics. This is a feature 
that has already been used in Ref.~\cite{Leonhardt:2019fua} to constrain the equation of state of symmetric nuclear matter. 

The only free parameter in our study is the value of the running gauge coupling $g_\mathrm{s}$ at the initial RG scale~$\Lambda$.  
This value is adjusted at the UV scale $\Lambda = 10\, \text{GeV}$ to obtain a critical temperature $T_\mathrm{cr} (\mu = 0) \equiv T_0 = 132 \, \text{MeV}$ 
at zero quark chemical potential, being the value of the chiral phase transition temperature at~$\mu=0$ 
found in very recent lattice QCD studies~\cite{Ding:2019prx}.\footnote{The lattice QCD 
study presented in Ref.~\cite{Ding:2019prx} considering two degenerate massless quarks and a physical strange quark 
mass finds the chiral phase transition $T_\mathrm{cr} = 132^{+3}_{-6}\, \text{MeV}$ at zero quark chemical potential.}
In Sec.~\ref{sec:fpps}, we shall use the scale $T_0$ as a reference scale, i.e., we ``measure'' all physical observables in units of this scale. 
The chosen value for~$\Lambda$ ensures that $T / \Lambda \ll 1$ and $\mu / \Lambda \ll 1$ 
for the range of external parameters considered in the present work 
and therefore allows us to avoid cutoff artefacts (and to reduce regularization-scheme dependences)~\cite{Braun:2018svj}.

In order to obtain the critical temperature $T_\mathrm{cr} (\mu = 0)= 132 \, \text{MeV}$ at~$\mu=0$, 
we tune the running gauge coupling at the initial scale~$\Lambda$. For example, for our study with 
a gauge coupling with two quark flavors, this amounts to choosing 
$\alpha_\mathrm{s}(\Lambda = 10 \, \text{GeV}) = g_{\rm s}^2/(4\pi) = 0.2137$. 
Evolved to the $Z$-boson mass scale $M_Z = 91.19\,\text{GeV}$, the value of the gauge coupling fixed in this way is 
then about 6\% greater than the experimental results~\cite{Tanabashi:2018oca}. 
The necessity of a larger initial value for the running gauge coupling in order to trigger criticality in the quark 
sector has been observed before in studies employing approximations of the present type, see, e.g., Ref.~\cite{Braun:2014ata}. 
Only most advanced RG truncations with a very accurate treatment of momentum structures do not require such 
an ``enhancement'', see, e.g., Refs.~\cite{Mitter:2014wpa,Cyrol:2017ewj}. The latter is of particular importance for 
a quantitative description of low-energy observables within the symmetry-broken regime. For our present work 
aiming at a study of the onset of spontaneous symmetry breaking and an analysis of 
symmetry breaking patterns, we expect that this 
plays a secondary role.

\section{Phase diagram}\label{sec:fpps}
\subsection{Symmetry breaking patterns}

Let us now study the phase diagram in the plane spanned by the temperature and the quark chemical potential 
for the $\UAone$-symmetric case, i.e., with all four-quark couplings set to zero at the initial RG scale~$\Lambda$. 
Recall that we have defined the critical temperature~$T_\mathrm{cr}(\mu)$ at a given value of the quark chemical potential 
to be the smallest temperature for which all four-quark couplings still remain 
finite in the IR limit $k\to 0$~\cite{Braun:2005uj, Braun:2006jd,Braun:2011pp,Braun:2017srn,Braun:2018bik}. 
Keep also in mind that a divergence in the flow of one four-quark coupling at a critical scale~$k_\mathrm{cr}(T, \mu)$ entails 
corresponding divergences in all other channels as well. 
However, as discussed in detail in Refs.~\cite{Braun:2017srn,Braun:2018bik}, 
the four-quark couplings in general develop distinct relative strengths and 
it is possible to identify a dominant four-quark channel, i.e., the modulus of the associated 
coupling is (significantly) greater than the absolute values of all the other four-quark couplings, see our general discussion above.
\begin{figure}[t]
\centering
\includegraphics[width=\linewidth]{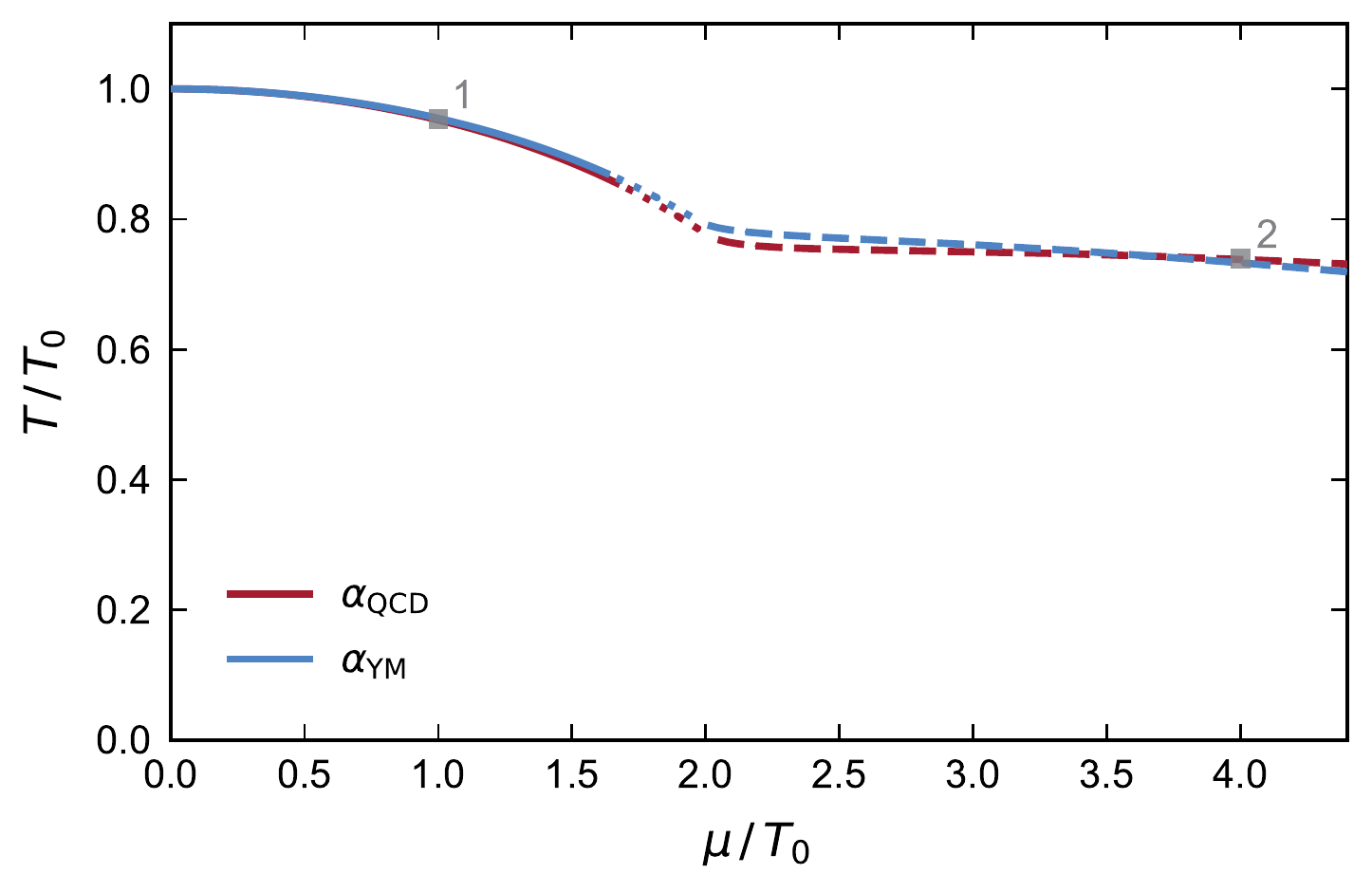}
\caption{Phase boundary associated with the spontaneous breakdown of at least one of the fundamental symmetries 
of QCD as accessible by our \Fierz-complete ansatz (red and blue lines).
The red line has been obtained by employing a strong coupling with two massless quark flavors. To illustrate the 
effect of quark contributions to the strong coupling, we also
show the phase boundary as obtained when the strong coupling from pure {\it Yang}-{\it Mills} theory is employed,
see main text for details.
The gray boxes labeled ``$1$'' and ``$2$'' specify two exemplarily points for which the RG flows of the four-quark couplings 
are shown in Fig.~\ref{fig:GaugeRGflow}.
}
\label{fig:GaugePhaseDiagram}
\end{figure}

As a first non-trivial result, we observe a dominance of the scalar-pseudoscalar coupling in the vacuum limit ($T=\mu=0$) 
with the couplings diverging at a symmetry breaking scale $k_\mathrm{cr}(\mu = 0)/ T_0 \approx 2.62$. 
Close to the symmetry breaking scale, we find that the modulus of the scalar-pseudoscalar coupling is 
at least two times greater than the modulus of all other couplings, suggesting that the QCD ground state is governed by 
chiral symmetry breaking in the vacuum limit.  
At first glance, this is very similar to the situation observed in NJL- and quark-meson-type studies. 
However, the crucial difference is given by the initial conditions for the four-quark couplings. In the present case, 
we can exclude that the dominance of the scalar-pseudoscalar channel may have been triggered by the choice 
for the initial conditions of the four-quark couplings since they are set to zero at the initial scale and are therefore only generated
dynamically. 
Thus, the dominance of the scalar-pseudoscalar coupling indicating chiral symmetry breaking is 
solely generated by quark-gluon dynamics.
This is in line with earlier RG flow studies of QCD in the vacuum limit~\cite{Braun:2006zz,Mitter:2014wpa} and 
it has indeed been observed that this dominance entails 
the formation of 
a corresponding chiral condensate~\cite{Mitter:2014wpa,Cyrol:2017ewj} governing the 
low-energy dynamics.

\begin{figure*}[t]
\centering
\includegraphics[width=0.83\linewidth]{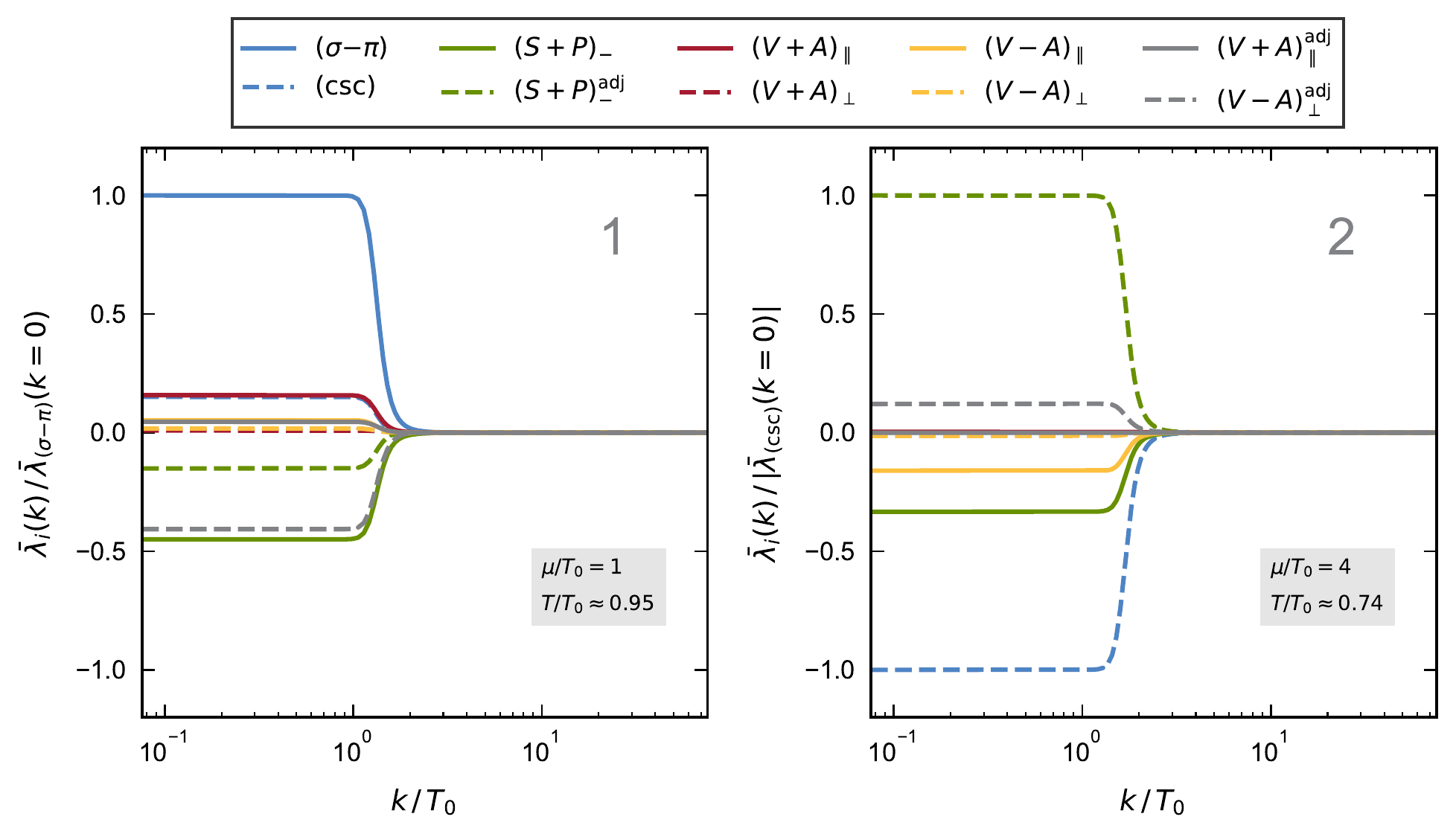}
\caption{Scale dependence of the various renormalized (dimensionful) four-quark couplings at $\mu / T_0 = 1$ 
and $T / T_0 \approx 0.95$ (left panel), as well as at $\mu / T_0 = 4.0$ and $T / T_0 \approx 0.74$ (right panel). These two parameter
sets correspond to the little gray boxes in the phase diagram shown in Fig.~\ref{fig:GaugePhaseDiagram}. 
According to the sum rules~\eqref{eq:SUA1} and~\eqref{eq:SUA2}, 
the intact $U_\mathrm{A}(1)$ symmetry implies $\bar \lambda _{\csc} = - \bar \lambda_{(S+P)_-^\mathrm{adj}}$, 
in agreement with our numerical results.
}
\label{fig:GaugeRGflow}
\end{figure*}

Increasing the temperature at vanishing quark chemical potential, 
the dominance of the scalar-pseudoscalar interaction channel persists even up to high temperatures 
beyond the critical temperature.
The red line in Fig.~\ref{fig:GaugePhaseDiagram} depicts the critical temperature as a function of the 
quark chemical potential which has been obtained with the running gauge coupling as computed 
in Refs.~\cite{Braun:2005uj,Braun:2006jd} 
for $\Nf=2$, here denoted by $\alpha_\mathrm{QCD}$. 
Following the phase boundary from small to large chemical potential, we first 
observe that the dominance of the scalar-pseudoscalar interaction channel 
persists up to $\mu / T_0 \approx 1.7$ as indicated by the red solid line. 
To illustrate the relative strengths of the various four-quark couplings in this regime, we show 
the scale dependence of the (dimensionful) renormalized couplings at $\mu / T_0 =1.0$ for~$T / T_0 \approx 0.95\gtrsim T_\mathrm{cr}(\mu)/T_0$ (i.e., 
right above the phase transition) in the left panel 
of Fig.~\ref{fig:GaugeRGflow}.  
This point is indicated in the phase diagram by the little gray box labeled~``$1$''. 
For illustrational purposes, 
the various couplings are normalized by the value of the scalar-pseudoscalar coupling $\bar \lambda_{\sigmapi}$ at $k = 0$. 
We readily 
observe that the dynamics is clearly dominated by the scalar-pseudoscalar coupling. In fact, its modulus is at least two times greater 
than the modulus of all other couplings. According to our line of argument, 
this dominance indicates that in this regime the phase boundary continues to be 
governed by chiral symmetry breaking.

Following the phase transition line, we then encounter that the exclusive dominance of the scalar-pseudoscalar 
channel goes away and 
a small region from approximately $\mu / T_0 \approx 1.7$ to $\mu / T_0 = \mu_\chi / T_0 \approx 2.0$ opens up 
(as depicted by the red dotted line in Fig.~\ref{fig:GaugePhaseDiagram}). In this region, 
we observe that the scalar-pseudoscalar channel, the CSC channel, 
as well as the $(S+P)_-^\mathrm{adj}$-, $(V+A)_\parallel^\mathrm{adj}$- and $(V-A)_\perp^\mathrm{adj}$-channel 
are most dominant, meaning that these channels are significantly greater compared to 
the remaining interaction channels. Such a  region of ``mixed'' dominances might potentially indicate a metastable or 
mixed phase~\cite{Roscher:2019omd}. It is noteworthy that four out of this set of five dominant channels are adjoint interaction 
channels which may point to a non-trivial color-structure of the ground state in this regime. 
However, the appearance of 
this regime may also very well be a consequence of the $\UAone$-preserving initial conditions 
as the resulting $\UAone$-symmetric RG flow affects the development of dominances. 
Indeed, the ``entanglement" of several equally strong four-quark couplings is lifted 
by taking into account $\UAone$-violating effects, see Subsec.~\ref{sec:GaugeUA1}.  

For $\mu/ T_0 \gtrsim 2.0$, depicted by the red dashed line in Fig.~\ref{fig:GaugePhaseDiagram}, 
we eventually observe a clear dominance of the CSC channel indicating the 
emergence of a diquark condensate~$\delta^{a}$. This dominance is again illustrated by the scale dependence of the 
couplings in this region shown in the right panel of Fig.~\ref{fig:GaugeRGflow}. There, we show 
the RG flow for $\mu / T_0 = 4.0$ and $T / T_0 \approx 0.74 \gtrsim T_\mathrm{cr}(\mu)/T_0$ (i.e., 
right above the phase transition). In the phase diagram, the corresponding point is depicted by 
the little gray box labeled ``$2$'' in Fig.~\ref{fig:GaugeRGflow}. 
Note that, in the right panel of Fig.~\ref{fig:GaugeRGflow}, 
the couplings are now normalized by the modulus of the dominant CSC coupling, $|\bar \lambda _{\csc} |$ at $k=0$. 
We observe that the modulus of any other four-quark coupling is at most less than half the value of the CSC coupling, 
except for the $(S+P)_-^\mathrm{adj}$ coupling. The latter  
assumes the same value in the IR as the CSC coupling, 
only with opposite sign. The reason for this behavior is that the boundary conditions with all four-quark couplings 
initially set to zero at the UV scale~$\Lambda$ leaves the axial $U_\mathrm{A}(1)$ symmetry intact, as already mentioned above. 
Since the RG flow preserves the symmetries of the initial effective average action, 
the sum rules~\eqref{eq:SUA1} and~\eqref{eq:SUA2} are exactly fulfilled at all scales~$k$, 
with the first sum rule implying $\bar \lambda _{\csc} = - \bar \lambda_{(S+P)_-^\mathrm{adj}}$. 
In fact, the sum rules show that two of the 10~four-quark couplings of our \Fierz-complete basis are not independent 
in the $U_\mathrm{A}(1)$-symmetric case. 
We shall discuss the effect of $\UAone$-violating initial discussions in Subsec.~\ref{sec:GaugeUA1}. 

\begin{figure}[t]
\centering
\includegraphics[width=\linewidth]{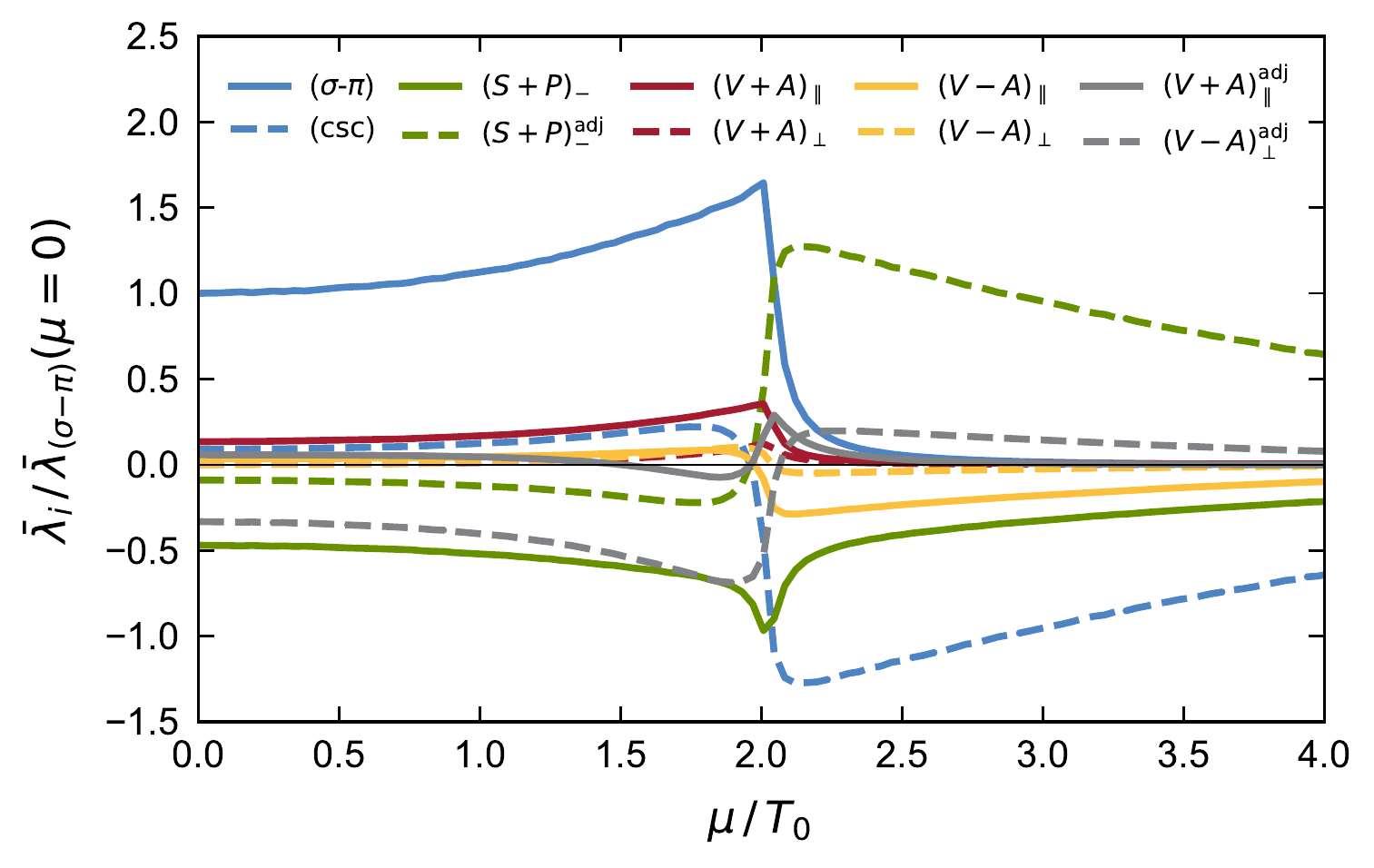}
\caption{(Dimensionful) renormalized four-quark couplings at $k=0$ 
as functions of the quark chemical potential for temperatures $(T-T_\mathrm{cr}(\mu))/T_0 \approx 0.004$ (i.e., slightly 
above the respective critical temperature $T_\mathrm{cr}$), 
illustrating the ``hierarchy'' of the four-quark couplings in terms of their relative strength 
along the phase boundary (i.e., along the red line in Fig.~\ref{fig:GaugePhaseDiagram}). For convenience, 
the values of the four-quark couplings are normalized by the value of the 
scalar-pseudoscalar coupling $\bar \lambda_{\sigmapi}$ for~$\mu=0$ at $k=0$.
}
\label{fig:DominancePattern}
\end{figure}

In Fig.~\ref{fig:DominancePattern}, we show the dominance pattern of the four-quark couplings 
along the finite-temperature phase boundary presented in Fig.~\ref{fig:GaugePhaseDiagram} (as obtained with 
the two-flavor coupling~$\alpha_{\rm QCD}$). 
The values of the dimensionful renormalized couplings are shown for $k\to 0$ as functions of the 
quark chemical potential for temperatures right above the critical temperature $T_\mathrm{cr}(\mu)$, where we have 
normalized the couplings by the scalar-pseudoscalar coupling $\bar \lambda_{\sigmapi}$ for~$\mu=0$ at $k=0$ for convenience.
As already mentioned above, we first observe a clear dominance of the scalar-pseudoscalar interaction channel 
for $\mu / T_0 \lesssim 1.7$, followed by a change of the dominance pattern to a 
dominance of the CSC coupling for $\mu/T_0\gtrsim 2$. In the region of CSC dominance, the intact $U_\mathrm{A}(1)$ symmetry is 
again encoded by the fact that the values of the CSC and the $(S+P)_-^\mathrm{adj}$-coupling are identical, up to their signs. 
The dominance of the CSC coupling for $\mu / T_0 \gtrsim 2$ is as clearly visible  
as the dominance of the scalar-pseudoscalar coupling for $\mu / T_0 \lesssim 1.7$. In fact, the 
ratio of the modulus of the second largest coupling and the largest coupling is even 
smaller in this high-density regime. 
Note that, in Fig.~\ref{fig:DominancePattern}, the values of the four-quark couplings are extracted 
at a temperature~$T$ slightly above the 
critical temperature~$T_\mathrm{cr}(\mu)$ where $(T-T_\mathrm{cr}(\mu)) / T_0 \approx 0.004$.
By this, we ensure that the RG flow is fully located in the symmetric regime and the flow can be followed down 
to~$k=0$.
However, owing to this small distance to the phase transition line, the aforementioned region 
with ``mixed'' dominances is not fully resolved in this figure.

The distinct dominance and evident ``hierarchy'' among the four-quark self-interactions in the low- and high-density 
regime is nicely illustrated in 
Fig.~\ref{fig:DominancePattern}. In view of this result, it is tempting to speculate whether 
such a change in the ``hierarchy'' points to the existence of a nearby tricritical point in the phase diagram. 
Yet, this aspect is speculative as the presently employed approximations do not allow a definite answer to this question. 
However, we emphasize that the change in the ``hierarchy'' from a dominance of the 
scalar-pseudoscalar coupling to a dominance of the CSC coupling at $\mu_\chi / T_0 \approx 2.0$ is a 
non-trivial outcome of our study as it is completely determined by the dynamics of the system itself. 
The four-quark couplings are initially set to zero at the UV scale~$\Lambda$ and are dynamically generated by 
quark-gluon dynamics in the RG flow. Thus, the dynamics is not affected by any kind of fine-tuning of the boundary 
conditions of the four-quark couplings which would potentially favor particular channels. 

At this point, let us once more bring to attention that the dominance of a four-quark coupling 
only indicates the onset of the formation of an associated condensate. It does neither guarantee the actual formation (e.g., 
IR fluctuations could restore the associated symmetries) nor does it strictly exclude the possible 
formation of other condensates associated with subdominant couplings.
Our analysis based on the dominance pattern of the four-quark couplings must therefore be taken with some care, 
see also our discussion in our preceding works~\cite{Braun:2017srn,Braun:2018bik}. Still, in the context of condensed-matter 
physics, the appearance of a clear dominance of a given channel 
has been found to be a precursor of the formation of a corresponding condensate~\cite{Roscher:2019omd}.  
\begin{figure*}[t]
\centering
\includegraphics[width=0.83\linewidth]{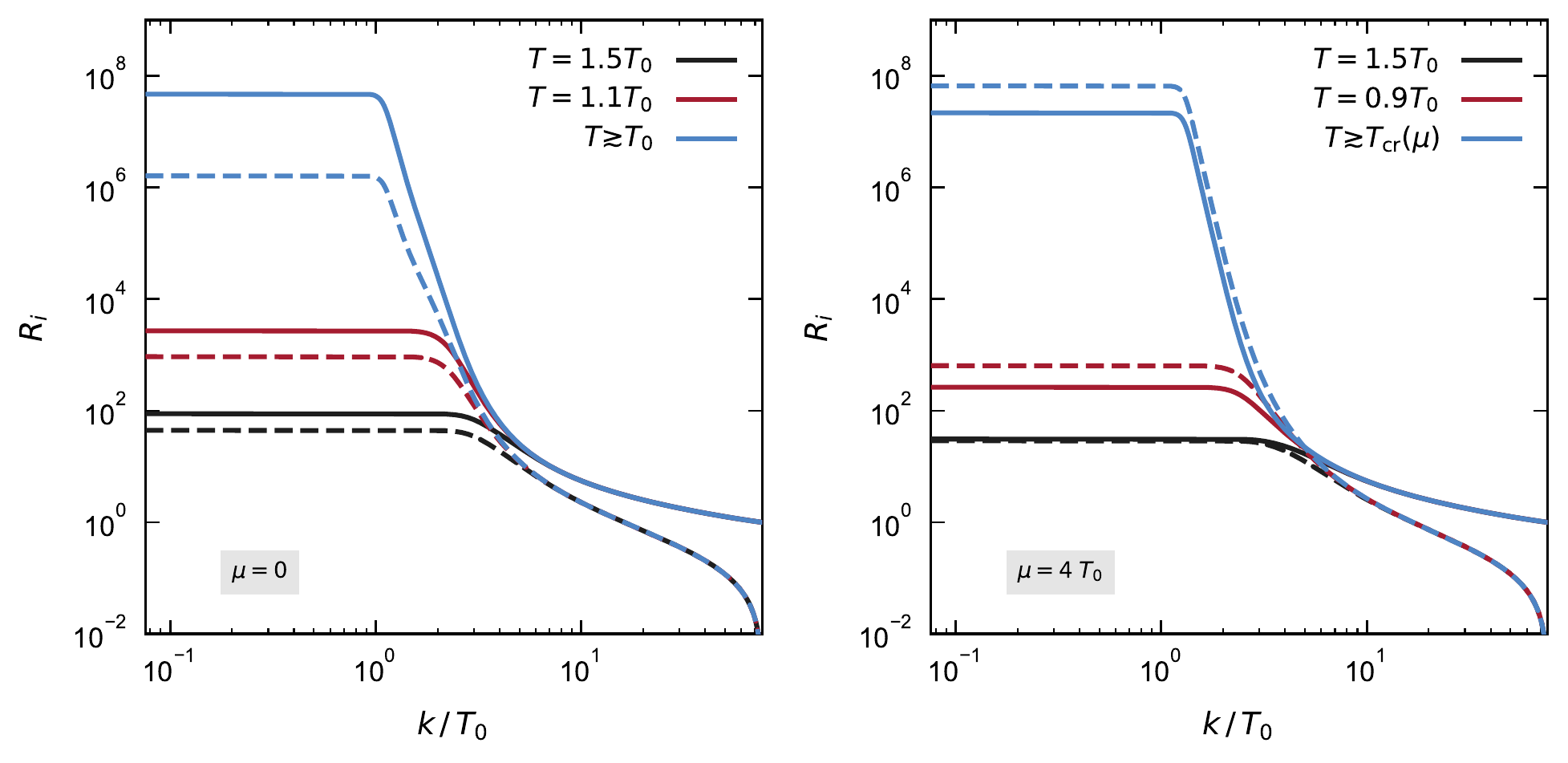}
\caption{Illustration of the scale dependence of the explicit $\UAone$ breaking as measured by the 
normalized sum rules~$R_1$ (dashed lines) and $R_2$ (solid lines) at~$\mu=0$ (left panel) and at $\mu/T_0=4.0$ (right panel) for three 
values of the temperature for each of the two cases. 
The strength of the initial explicit $\UAone$ symmetry breaking is controlled by the value of the (dimensionless) 
renormalized coupling of the $(S+P)_-$-channel (``\textit{'t Hooft} channel") at the UV scale~$\Lambda$, 
exemplarily chosen to be~$\lambda_{(S+P)_-}^\mathrm{(UV)}=1.0$ in this figure.
}
\label{fig:SumRules}
\end{figure*}

In Fig.~\ref{fig:GaugePhaseDiagram}, 
we also included results for the finite-temperature phase boundary from 
a computation where we have used a running gauge coupling as obtained in 
YM~theory (depicted by the blue line labeled $\alpha_\mathrm{YM}$).
The phase boundary as well as the dominances agree almost perfectly with the results 
from the computation using the running coupling~$\alpha_\mathrm{QCD}$. 
To be specific, we observe a dominance of the scalar-pseudoscalar coupling at small quark chemical potential, 
a regime characterized by a ``mixed'' dominance pattern between $1.7 \lesssim \mu / T_0 \lesssim 2.0$, 
and finally a clear dominance of the CSC coupling at large quark chemical potential.
This is noteworthy since the YM coupling~$\alpha_\mathrm{YM}$ 
grows more rapidly than 
the QCD coupling $\alpha_\mathrm{QCD}$ when the scale~$k$ is lowered. 
However, the effect of this difference in the scale dependence is also 
to a certain extent compensated by our scale fixing procedure. 
The initial value of the YM gauge coupling has to be chosen smaller than 
for the~$\alpha_\mathrm{QCD}$ coupling in order to obtain the same 
critical temperature $T_0\equiv T_\mathrm{cr}(\mu = 0) = 132\, \text{MeV}$ at zero quark chemical potential. 
Nevertheless, this observation may be viewed as an indication that the ``hierarchy'' of the various couplings in terms of their relative strength is 
predominantly determined by the dynamics within the quark sector whereas the gauge sector as associated with the details of the 
running coupling is mostly required to bring the quark sector close to criticality in the first place. Once the gauge sector have rendered the four-quark couplings
sufficiently large, they become relevant operators in the RG flow and the quarks start to develop their ``own effective dynamics". Then,  
the details of the gauge sector play a subleading role, at least in the present approximation. Loosely speaking, one may therefore
state that we are basically left with an NJL-type model 
once the four-quark couplings
have been rendered sufficiently large by the underlying quark-gluon 
dynamics.

\subsection{Role of $U_{\mathrm A}(1)$ symmetry}\label{sec:GaugeUA1}
The initial conditions of the RG flow chosen so far leave the axial $U_\mathrm{A}(1)$ symmetry 
intact. All couplings of the four-quark self-interactions are set to zero at the UV scale~$\Lambda$ 
and are solely generated dynamically by quark-gluon interactions in the RG flow. 
In, e.g., Refs.~\cite{Braun:2005uj,Braun:2006jd}, $U_\mathrm{A}(1)$-violating channels 
have been omitted based on the assumption that they become relevant 
only in the low-energy regime governed by spontaneous symmetry breaking. 
Our \Fierz-complete basis $\mathcal B$ composed of the 10~four-quark interaction 
channels is effectively reduced to eight interaction channels 
in case of the $\UAone$ symmetry being intact~\cite{Braun:2018bik}. 
Recall that the sum rules~\eqref{eq:SUA1} and~\eqref{eq:SUA2} 
imply that two of the couplings associated with the four $\UAone$-violating interaction channels of our basis $\mathcal B$ 
are not independent. 
These sum rules are exactly fulfilled at all scales in the symmetric phase and for all $k > k_\mathrm{cr}$ in the phase 
governed by spontaneous symmetry breaking. In the latter case, the $\UAone$ symmetry may potentially still 
be broken spontaneously below the symmetry breaking scale~$k_\mathrm{cr}$. However, this cannot be resolved within
our present approximation. 

In our preceding study of the phase structure of the NJL model and the 
role of \Fierz completeness~\cite{Braun:2018bik}, 
we have observed that $\UAone$ symmetry breaking affects the dominances 
of the four-quark couplings in terms of their relative strength along the finite-temperature phase 
boundary. In particular, we have found explicit~$\UAone$ symmetry breaking to be important 
for the formation of the conventional CSC ground state at intermediate and large values of the chemical potential. 
In this work, the four-quark couplings are now dynamically generated by quark-gluon dynamics.
Following the critical temperature~$T_\mathrm{cr}(\mu)$ as a function of the quark chemical potential, we observe 
two regions characterized by different distinct ``hierarchies'' of the four-quark couplings which are 
remarkably robust against a variation of the running gauge coupling, see 
Fig.~\ref{fig:GaugePhaseDiagram} for the $\UAone$-symmetric case. 
The scalar-pseudoscalar interaction channel dominates the dynamics at small quark chemical potential, 
signaling the formation of the chiral condensate, whereas at higher quark chemical potential the 
dominance of the conventional CSC coupling suggests the formation of a diquark condensate. 
The latter is observed in spite of the intact $\UAone$ symmetry in our considerations thus far.
Only for $1.7 \lesssim \mu / T_0 \lesssim 2.0$, we observe a regime 
which is characterized by several equally strong four-quark interaction channels, see our discussion in the 
previous subsection. 
Moreover, the dominance of the CSC coupling is always accompanied by an equally strong $(S+P)_-^\mathrm{adj}$-channel 
as a direct consequence of the intact $\UAone$ symmetry: The sum rule~\eqref{eq:SUA1} ties the 
modulus of the CSC coupling to the modulus of the $(S+P)_-^\mathrm{adj}$ coupling. 

In order to probe the role of explicit $\UAone$-symmetry breaking in our present study,
 we now analyze the RG flow for $\UAone$-violating boundary conditions for the four-quark couplings. 
The strength of $\UAone$ breaking is effectively controlled by the initial value of the $(S+P)_-$ coupling 
since the associated four-quark interaction channel is directly related to the so-called \textit{'t Hooft} 
determinant~\eqref{eq:det}, see Refs.~\cite{tHooft:1976rip,tHooft:1976snw,Shifman:1979uw,Shuryak:1981ff,
Klevansky:1992qe,Schafer:1996wv,Pawlowski:1996ch}. In the following, we therefore vary only 
the initial condition of this coupling but still set 
the initial values of the other four-quark couplings to zero at the UV scale~$\Lambda$.  
For a given initial value of the $(S+P)_-$ coupling, we then adjust the UV value of the gauge coupling~$\gs(\Lambda)$ 
such that the value of the critical temperature at zero chemical potential remains unchanged, $T_\mathrm{cr}(\mu = 0) = 132\, \text{MeV}$. This 
ensures comparability between our results for different initial conditions for the $(S+P)_-$ coupling.
\begin{figure}[t]
\centering
\includegraphics[width=\linewidth]{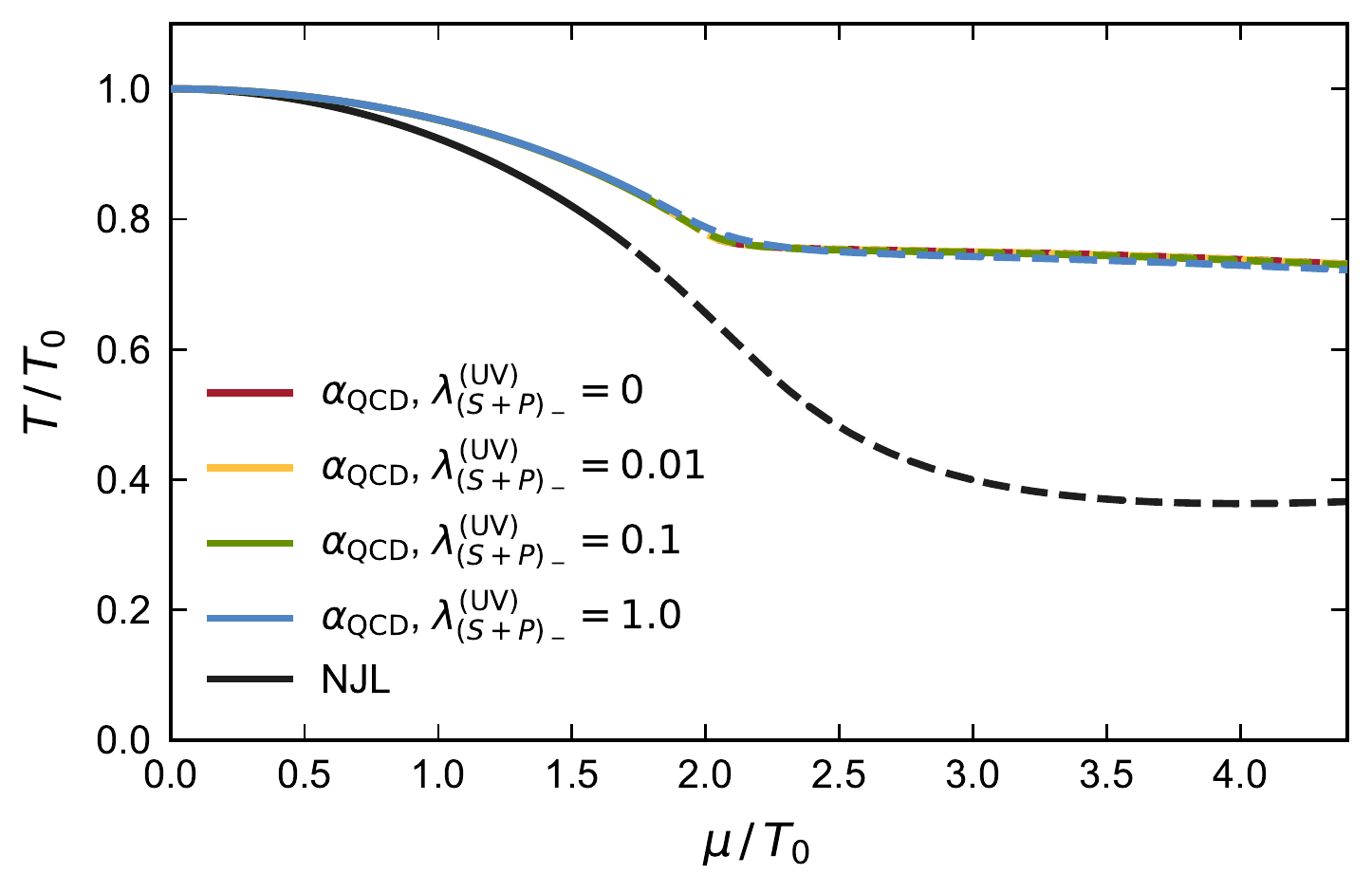}
\caption{Phase boundary associated with the spontaneous breakdown of at least one of the fundamental symmetries 
of QCD as accessible by our \Fierz-complete ansatz now under the influence 
of explicit $\UAone$ symmetry breaking, 
in comparison to the phase boundary resulting from a corresponding \Fierz-complete NJL model study (black line), see 
also Ref.~\cite{Braun:2018bik}. 
All results (except for those from the NJL model study) have been obtained by employing a strong coupling with two massless quark flavors. 
The strength of the initial explicit $\UAone$ symmetry breaking is controlled by the value of the (dimensionless) 
renormalized coupling of the $(S+P)_-$ channel (``\textit{'t Hooft} channel") at the UV scale~$\Lambda$. 
The values of all other four-quark couplings have been chosen to be initially zero. The phase boundary is shown for $\UAone$-symmetric boundary conditions 
as well as for $\UAone$-violating initial conditions with~$\lambda_{(S+P)_-}^\mathrm{(UV)}=0.01,\, 0.1,\, 1.0\,$. 
A dominance of the scalar-pseudoscalar interaction channel is depicted by solid lines and a dominance of the CSC channel by 
dashed lines. The case of ``mixed'' dominances occurring for $\UAone$-symmetric boundary conditions is indicated by a dotted line 
although hardly visible on the scale of the plot. 
}
\label{fig:GaugePhaseDiagramUA1}
\end{figure}

As also done in Refs.~\cite{Braun:2017srn,Braun:2018bik}, we begin 
by analyzing the fate of the $\UAone$ symmetry at finite temperature and quark chemical potential 
with the aid of the sum rules~\eqref{eq:SUA1} and~\eqref{eq:SUA2}.
To this end, we first normalize the two sum rules as follows:
\be
1 =\left( R_1 + R_2 \right)\big|_{k=\Lambda}\,,
\ee
where
\be
R_i = \mathcal{N} \left| \mathcal{S}_{U_\mathrm{A}(1)}^{(i)} \right|\,.
\label{eq:NJL2Rdef}
\ee
The quantities $\mathcal{S}_{U_\mathrm{A}(1)}^{(1)}$ 
and~$\mathcal{S}_{U_\mathrm{A}(1)}^{(2)}$ are defined in 
Eqs.~\eqref{eq:SUA1} and~\eqref{eq:SUA2}, respectively. 
A strong deviation of the sum rules from zero indicate strong effective~$\UAone$ breaking.
In Fig.~\ref{fig:SumRules}, we show the scale dependence of the sum rules 
at two characteristic values of the chemical potential for  three 
values of the temperature for each of the two cases. 
Interestingly, we observe the exact same qualitative behavior as found in Ref.~\cite{Braun:2018bik} for the NJL model.  
At small quark chemical potential close to the critical temperature~$T_\mathrm{cr}(\mu)$, $\UAone$ 
breaking is driven by the scalar-pseudoscalar channel associated with pion dynamics
and becomes stronger toward the IR as indicated by increasing values of~$R_1$ and~$R_2$, with~$R_2 \gg R_1$.
At large chemical potential, we find that the strength of $\UAone$-symmetry breaking becomes also stronger 
as the phase boundary is approached from above, but now driven by the dynamics of diquark degrees of freedom 
as associated with the CSC channel. As a consequence, $R_1$ and $R_2$ are of the 
same order of magnitude since both depend on the CSC coupling, see Eqs.~\eqref{eq:SUA1} and~\eqref{eq:SUA2}. 
In either case, for increasing temperature, $\UAone$ breaking as measured by the sum rules 
remains more and more on its initial level as determined by the $\UAone$-violating boundary conditions in the UV regime 
since quark fluctuations become more and more thermally suppressed.

Let us now compare the phase diagram as obtained with the $\UAone$-symmetric 
initial conditions employed in the previous subsection, i.e., with all four-quark couplings initially set to zero, 
to the phase diagrams resulting from $\UAone$-violating initial conditions. 
The strength of the explicit $\UAone$ breaking at the initial UV scale~$\Lambda$ is 
controlled by the value of the (dimensionless) renormalized coupling of the $(S+P)_-$ channel (``\textit{'t Hooft} channel")
which we choose to assume the values $\lambda_{(S+P)_-}^\mathrm{(UV)}=0.01,\, 0.1,\, 1.0\,$. 
In the following, we shall only present results from computations using the running gauge coupling for two massless quark flavors. 
As in the $\UAone$-symmetric case, the dependence on our specific choice 
for the coupling is found to be very mild anyhow.
In Fig.~\ref{fig:GaugePhaseDiagramUA1}, the various phase diagrams as obtained with these three  
different choices for the boundary conditions are shown. It is remarkable how little the critical phase temperature 
as a function of the quark chemical potential is affected by the strength of the initial explicit 
breaking of the $\UAone$ symmetry, although the strength in terms of the initial value of the $(S+P)_-$ coupling is 
varied over three orders of magnitude.\footnote{Note that the considered range of 
values for the $(S+P)_-$ coupling is consistent with the size of the value that is expected 
from a direct computation of this quantity at a given scale~$\Lambda$~\cite{Pawlowski:1996ch}.} 
Across the entire range of chemical potentials 
shown in this figure, the variation of the critical temperature for any given value of the 
chemical potential is less than $2\%$.\footnote{For all initial 
conditions of the computations shown in Fig.~\ref{fig:GaugePhaseDiagramUA1}, 
the symmetry breaking scale in the vacuum limit remains at approximately the same value, $k_\mathrm{cr}/T_0 \approx 2.6$. 
Still, a direct quantitative comparison of the phase boundaries has to be taken with some care 
as the different computations do not necessarily lead to the same values of low-energy observables.}
In all cases, we observe a dominance of the scalar-pseudoscalar coupling at small quark chemical potential, 
depicted by the solid lines in Fig.~\ref{fig:GaugePhaseDiagramUA1}.
The regime of ``mixed'' dominances for chemical potentials between $1.7 \lesssim \mu / T_0 \lesssim 2.0$ 
appearing in case of $\UAone$-symmetric initial conditions vanishes for {\it all} considered $\UAone$-violating 
initial conditions. Choosing $0.01 \leq \lambda_{(S+P)_-}^\mathrm{(UV)} \leq 1.0$ for the initial value 
of the coupling associated with the $\UAone$-violating $(S+P)_-$ channel, 
we indeed find that the dominance changes directly from the scalar-pseudoscalar channel to the CSC channel 
within the region $1.8 \lesssim \mu_\chi / T_0 \lesssim 2.0$, 
as indicated by the dashed lines in Fig.~\ref{fig:GaugePhaseDiagramUA1}. This is similar to 
what has been found in a \Fierz-complete NJL-model study~\cite{Braun:2018bik}.
\begin{figure}[t]
\centering
\includegraphics[width=\linewidth]{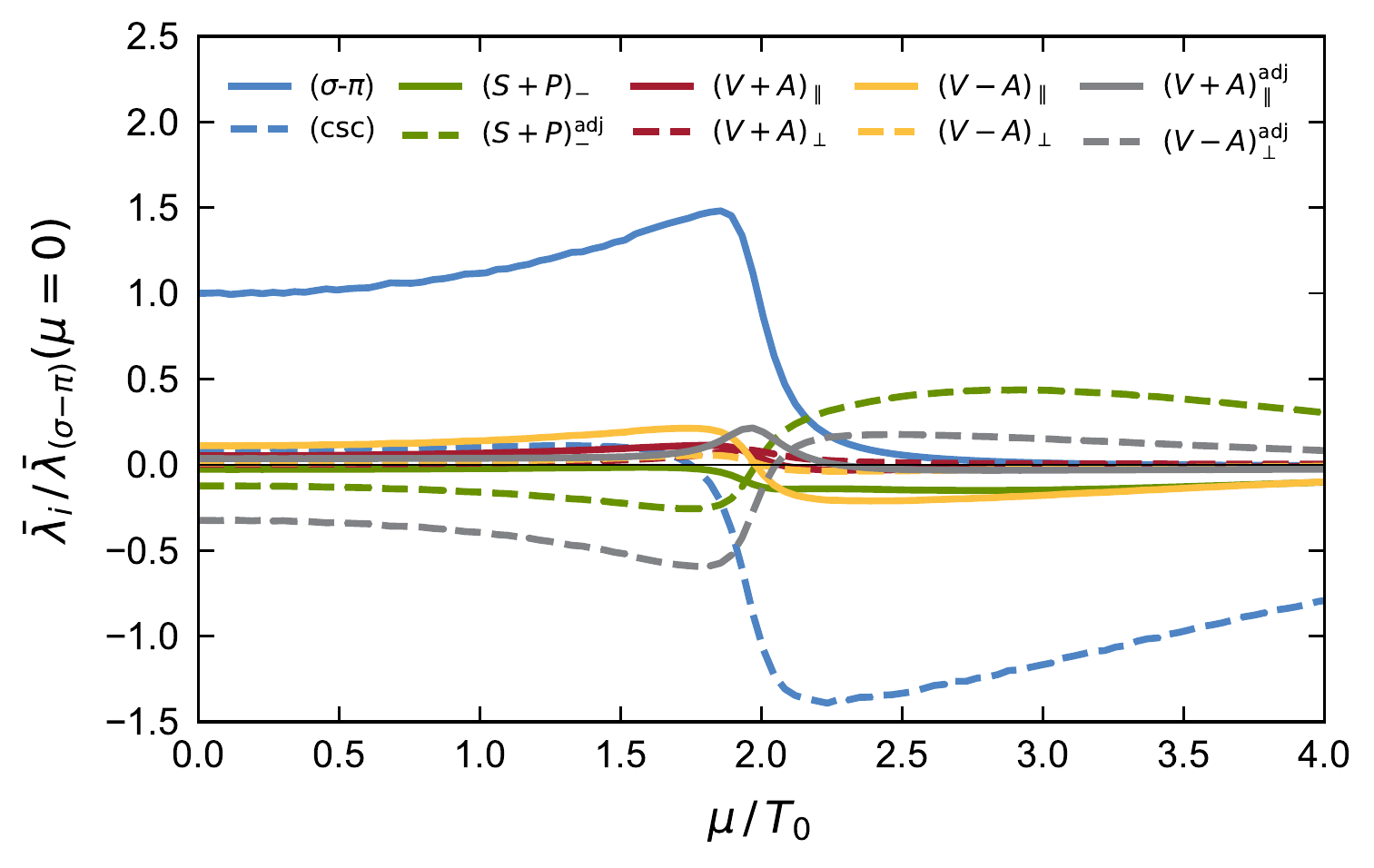}
\caption{``Hierarchy'' of the four-quark couplings in terms of their relative strength 
along the phase boundary (blue line in Fig.~\ref{fig:GaugePhaseDiagramUA1}). To obtain this figure, we have 
evaluated the renormalized four-quark couplings for $k\to 0$ as functions of the quark chemical potential for 
temperatures $(T-T_\mathrm{cr}(\mu))/T_0 \approx 0.002$ (i.e., slightly above the critical temperature~$T_\mathrm{cr}(\mu)$ 
for a given quark chemical potential).
The values of the four-quark couplings are shown for $\UAone$-violating initial conditions with $\lambda_{(S+P)_-}^\mathrm{(UV)}=1.0$ 
for the \textit{'t Hooft} coupling.
To normalize the couplings, we have used the scalar-pseudoscalar coupling for $k\to 0$ and zero quark chemical potential.
}
\label{fig:DominancePatternUA1}
\end{figure}

The ``hierarchy'' of the various four-quark couplings in terms of their relative strength 
along the phase boundary is shown in Fig.~\ref{fig:DominancePatternUA1}, exemplarily 
for the initial coupling $\lambda_{(S+P)_-}^\mathrm{(UV)}=1.0$. This figure shows again the values 
of the (dimensionful) renormalized couplings for $k\to 0$ as functions of the quark chemical potential for 
temperatures $(T-T_\mathrm{cr}(\mu)) / T_0 \approx 0.002$ (i.e., slightly above the critical temperature~$T_\mathrm{cr}(\mu)$ 
for a given quark chemical potential~$\mu$). The values are normalized by the scalar-pseudoscalar coupling $\bar \lambda_{\sigmapi}$ 
for~$k\to 0$ at zero chemical potential.
We again observe a clear dominance of the scalar-pseudoscalar interaction channel for $\mu / T_0 \lesssim 2.0$ 
and a clear dominance of the CSC interaction channel for larger values of the chemical potential. 
Compared to the case with $\UAone$-symmetric 
boundary conditions (see Fig.~\ref{fig:DominancePattern}), these dominances appear even more pronounced in the present case. 
In particular, the CSC channel is not accompanied anymore by an equally strong $(S+P)_-^\mathrm{adj}$ coupling. 
The latter now assumes considerably smaller values for $\mu / T_0 \gtrsim 2.0$ in comparison to the computation 
with intact $\UAone$ symmetry, whereas the CSC coupling assumes even slightly larger values. 
From this, we conclude that the breaking of the $U_A(1)$ symmetry plays an important role in 
``shaping the hierarchy'' of the four-quark interaction channels 
and thus in the formation of associated condensates as indicated by their dominances. 
This observation confirms the importance of explicit~$U_\mathrm{A}(1)$ breaking for the formation 
of the conventional CSC ground state at large chemical potential as already discussed 
in Ref.~\cite{Braun:2018bik} for NJL-type models.  In this respect, we also refer again to early seminal works on 
color superconductivity, see, e.g., 
Refs.~\cite{Alford:1997zt,Rapp:1997zu,Berges:1998rc,Son:1998uk,Pisarski:1999tv,Pisarski:1999bf,Schafer:1999jg,Brown:1999aq,Hong:1999fh,Evans:1999at}.
Noteworthy, we find that the change in the ``hierarchy'' from a dominance of the scalar-pseudoscalar coupling to a 
dominance of the CSC coupling at $\mu_\chi / T_0 \approx 2.0$ is remarkably insensitive to the initial strength of explicit $\UAone$ 
symmetry breaking as controlled by the initial coupling~$\lambda_{(S+P)_-}^\mathrm{(UV)}$ associated with the ``\textit{'t Hooft} channel". 
We emphasize that this change in the ``hierarchy'' of four-quark couplings is a 
non-trivial outcome completely determined by the dynamics of the system itself.

Let us finally compare our results with those from a \Fierz-complete NJL model study~\cite{Braun:2018bik}.
In Fig.~\ref{fig:GaugePhaseDiagramUA1}, we also show the finite-temperature phase boundary resulting from 
a \Fierz-complete NJL model study (black line). The corresponding flow equations have been presented in 
Ref.~\cite{Braun:2018bik}. 
In this computation, the initial scalar-pseudoscalar coupling at the UV scale $\Lambda / T_0 \approx 75.76$ 
has been tuned such that we obtain $T_\mathrm{cr}(\mu=0) = 132\, \text{MeV}$ for the critical temperature at~$\mu=0$. 
The remaining initial four-quark couplings are set to zero. 
Interestingly, we find that the finite-temperature phase boundary agrees well with the one obtained from our 
present study with a two-flavor running gauge coupling~$\alpha_\mathrm{QCD}$, at least for small quark chemical potential. 
For~$\mu / T_0 \gtrsim 0.5$, however, the two phase boundaries start to deviate from each other significantly.   
Indeed, at the largest quark chemical potential shown in Fig.~\ref{fig:GaugePhaseDiagramUA1}, $\mu / T_0 = 4.4$, 
the critical temperature resulting from the NJL model computation is $T_\mathrm{cr} / T_0 \approx 0.366$. 
In contrast to that, at the same quark chemical potential, 
the computation including dynamical gauge fields yields a critical temperature 
that exceeds the one from the NJL model study by a factor of two: 
$T_\mathrm{cr}(\mu  = 4.4\,T_0) / T_0 \approx 0.731$. This observation may have further phenomenological consequences. 
For example, in standard BCS theory, the critical temperature can be directly related to the size of the 
diquark gap~$\delta^a$ at zero temperature, i.e., $T_\mathrm{cr} \sim | \delta |$~\cite{Bailin:1983bm,Schmitt:2002sc,Altland:2006si}.  
Thus, the observed increase of the critical temperature in our results from the 
computation including dynamical gauge degrees of freedom 
may hint to a larger diquark gap at~$T=0$.

Recall that we use different initial conditions for the four-quark couplings in our NJL model study and our 
study with dynamical gauge fields.
This is required since different mechanisms are at play which drive the quark sector to criticality. 
To be specific, in our NJL model study, it is required to choose a sufficiently large initial scalar-pseudoscalar coupling to 
ensure that the RG flow diverges at a finite symmetry breaking scale~$k_\mathrm{cr}$ for sufficiently low temperatures, 
signaling the onset of spontaneous symmetry breaking. Apart from the value of the coupling 
associated with the \textit{'t Hooft} channel, the four-quark couplings in our QCD study are 
solely dynamically generated and the quark sector is driven to criticality by the gauge coupling becoming sufficiently large, 
see Refs.~\cite{Gies:2005as,Braun:2005uj,Braun:2006jd,Braun:2011pp} for a detailed discussion of the latter mechanism. 
Of course, this difference also affects the results at large quark chemical potential, although 
the gauge coupling plays a less prominent role in this regime 
as the dynamics is largely controlled by the appearance of a BCS-type instability. 
One may argue that the initial conditions chosen in case of the NJL model study 
actually favor the scalar-pseudoscalar coupling and 
do not sufficiently support the dynamics associated with the formation of a diquark condensate or other channels
which may become relevant at large chemical potential.
Indeed, the boundary conditions enforce that the dynamics are initially driven by the scalar-pseudoscalar self-interaction, 
at least over a wide range of RG scales. 
Still, at large chemical potential, the CSC channel is found to dominate the dynamics even in case of the NJL model. 
As discussed in detail 
in Refs.~\cite{Braun:2018bik}, this 
can be understood from an analysis of the fixed-point structure.
Our study taking into account gauge degrees of freedom comes without the requirement 
of an initial tuning of a specific four-quark coupling. 
It is therefore intriguing that the ``hierarchy'' of the various interaction channels 
in both computations changes at approximately the same quark chemical potential  
from the scalar-pseudoscalar coupling to the CSC coupling, see Fig.~\ref{fig:GaugePhaseDiagramUA1}.
This observation indicates that the ``hierarchy'' of the interaction channels in terms of their strength is 
determined to a large extent 
by the interplay of the various four-quark couplings themselves, see also our discussion in the previous subsection.

As a closing remark, we would like to add that the comparison of the different phase boundaries 
shown in Figs.~\ref{fig:GaugePhaseDiagram} and~\ref{fig:GaugePhaseDiagramUA1} have to be taken with some care. Although all  
computations yield approximately the same critical scale $k_\mathrm{cr} / T_0 \approx 2.6$ (symmetry-breaking scale) 
in the vacuum limit,
which sets the scale for low-energy observables~${\mathcal O}\sim k_\mathrm{cr}$~(see, e.g., Ref.~\cite{Braun:2009ns}), 
our present approximation does not allow to check whether 
the different studies indeed lead to the same values of low-energy observables in the IR limit. 
This potential issue complicates a comparison of our results for the phase boundary, as well as 
our subsequent comparison of the curvature of the phase 
boundary.

\subsection{Curvature of the phase boundary}
Finally, we briefly comment on the curvature~$\kappa$ of the finite-temperature phase boundary at small chemical potential:
\be
\kappa = -{T_0}\frac{{\rm d} T_{\rm cr}(\mu)}{{\rm d}\mu^2}\Bigg|_{\mu=0}\,.
\ee
The results for the curvature as obtained from various different studies are 
summarized in Tab.~\ref{tab:GaugeCurvature}. 
\begin{table}[tb]
\centering
\begin{tabular}{c | c}
\hline\hline
 Setting &  curvature $\kappa$ \\
 \hline
mean field (NJL, one channel)~\cite{Buballa:2003qv,Aoki:2017rjl} & 0.197\dots 0.200\\[0.1cm] 
fRG (NJL, \Fierz-complete) &  0.074\\[0.1cm]
fRG (QCD, $\UAone$-symmetric) & 0.046\\[0.1cm]
%fRG ($\alpha_\mathrm{YM}$) & 0.044\\[0.1cm]
fRG (QCD) & 0.046\\[0.1cm]
Lattice QCD~\cite{deForcrand:2002hgr,Allton:2003vx,Wu:2006su,Philipsen:2007rj} & $0.034 \dots 0.070$\\
 \hline\hline
\end{tabular}
\caption{
Curvature~$\kappa$ of the two-flavor finite-temperature phase boundary as obtained from 
different studies, see main text for details. Note that the curvature range for the mean-field studies reflects the 
difference between the chiral limit and the case of physical pion masses. The fRG results in this work have been 
obtained in the chiral limit.
}
\label{tab:GaugeCurvature} 
\end{table}

Compared to our \Fierz-complete NJL model study, we find the curvature to be significantly decreased in 
our study including dynamical gauge degrees of freedom. In fact, the curvature is reduced by approximately~$40\%$. 
The curvature of a standard one-channel NJL-model study in the mean-field approximation is even more than 
four times greater than in our present study with gauge degrees of freedom. 
This holds true for all settings with gauge degrees of freedom considered in this work,  
including computations using a {\it Yang}-{\it Mills} coupling, a strong coupling for the two-flavor 
case, as well as 
computations taking into account explicit $\UAone$ symmetry breaking (labelled ``fRG (QCD)" in Tab.~\ref{tab:GaugeCurvature}). 
Note that the results from our \Fierz-complete fRG 
calculations including gauge degrees of freedom listed in Tab.~\ref{tab:GaugeCurvature} have been obtained 
with the strong coupling for the two-flavor 
case. For the case with explicit $\UAone$ symmetry breaking, we have 
chosen $\lambda^\mathrm{(UV)}_{(S+P)_-}=1.0$ for the initial condition of the coupling associated with the {\it 't~Hooft} channel.
A summary of results for the curvature from lattice QCD calculations with two flavors
can also be found in Tab.~\ref{tab:GaugeCurvature}. We observe that the results from 
our \Fierz-complete studies taking into account gauge degrees of freedom 
are well in accordance with those from lattice QCD studies. 
We add that low-energy model studies indicate that the inclusion of IR fluctuation effects 
tend to further lower the value of the curvature~\cite{Schaefer:2004en,Braun:2011iz,Pawlowski:2014zaa,Aoki:2017rjl}. 
For the~$2+1$-flavor case, more recent lattice results for the curvature  
are also available, indicating $\kappa \sim 0.034\, \dots 0.21$~\cite{DElia:2018fjp}. 
The curvature found in a very recent $2+1$-flavor fRG-QCD study with physical masses is also in accord 
with the latter results~\cite{Fu:2019hdw}.
In any case, 
a direct comparison of our present results to those from $2+1$-flavor studies is only possible to a limited extent, if at all. 
Still, in general, it is reasonable to expect from our present study that the issue of \Fierz-incompleteness also affects the results 
for the curvature in the $2+1$-flavor case, as it does in the $2$-flavor case~(see Tab.~\ref{tab:GaugeCurvature}).
Indeed, it has also been observed in a \Fierz-incomplete two-channel study of a $2+1$-flavor NJL-type model that 
four-quark interaction channels other than the scalar-pseudoscalar channel can significantly impact the value of the curvature~\cite{Bratovic:2012qs}.

\section{Conclusions}\label{sec:conc}
In this work, we have analyzed the RG flow of 
four-quark interactions in the pointlike limit in a \Fierz-complete fashion starting from the classical QCD action in the UV limit.
Working in the chiral limit, the only parameter of our study in the $\UAone$-symmetric limit 
is given by the strong coupling~$\gs$ which we fixed at a large initial UV scale in the perturbative regime.

With this setting at hand, we found that the inclusion of gluodynamics 
leads to an increase of the critical temperatures at large quark chemical potential in comparison to the results 
from a corresponding \Fierz-complete NJL model study. 
Assuming that the critical temperature can be related to the size of the zero-temperature gap in a color superconducting 
phase of quark matter, our results therefore 
suggest that the diquark gap is likely to be greater than the one found in 
corresponding NJL model studies, at least within the range of chemical potentials considered in our present work. 

Toward the IR limit, the treatment of the four-quark interactions in the pointlike limit does of course 
not allow us to access the phase governed by spontaneous symmetry breaking. 
The introduction of mesonic auxiliary fields by means of a \HubbardS transformation or 
applying the more advanced technique of dynamical hadronization, see 
Refs.~\cite{Gies:2001nw,Gies:2002hq,Pawlowski:2005xe,Floerchinger:2009uf} and, e.g., 
Refs.~\cite{Braun:2008pi,Mitter:2014wpa,Braun:2014ata,Cyrol:2017ewj,Fu:2019hdw} for their application to QCD, 
would enable us to study the dynamics even within regimes governed by spontaneous symmetry breaking. 
In order to gain nevertheless some insight into the structure of the ground state emerging in case of spontaneous symmetry breaking, we 
followed the approach of our NJL model studies in Refs.~\cite{Braun:2017srn,Braun:2018bik} and 
analyzed the ``hierarchy'' of the four-quark couplings. 

Our RG analysis of the $\UAone$-symmetric case revealed 
a clear dominance of the scalar-pseudoscalar interaction channel 
associated with chiral symmetry breaking
at small quark chemical potential. Very importantly, this dominance is not triggered 
by a specific choice for the initial conditions of the four-quark couplings since all four-quark couplings are 
set to zero at the initial RG scale, i.e., they 
are solely gluon-induced in the RG flow. 
For $\mu / T_0 \gtrsim 2.0$, we then observe a change in the ``hierarchy''. In this regime, 
the CSC channel associated with the formation of the most conventional color superconducting condensate in two-flavor QCD 
now dominates the quark dynamics. 
We emphasize that the dominance pattern as a function of the quark chemical potential 
as well as the actual position~$\mu_{\chi}$, where the dominance 
pattern changes from a scalar-pseudoscalar dominance to a 
CSC dominance,  
are found to be a remarkably robust feature. 
The little influence of the considered different running gauge couplings 
may be viewed as an indication that the dominance pattern is largely determined within the quark sector. 
The gauge sector as associated with the details of the 
running coupling is mostly required to bring the quarks close to criticality. Once the four-quark couplings
have been rendered sufficiently large by the underlying quark-gluon dynamics, 
the quarks develop their ``own dynamics" and 
the details of the gauge sector start to play a subleading role, at least in the present approximation. At this point, 
loosely speaking, we are then basically left with an NJL-type model. 
Note again that the dynamics in our present study is not ``contaminated" by any kind of fine-tuning of the initial 
conditions for the four-quark couplings which would in general favor particular channels.
However, the analysis based on the ``hierarchy'' of the four-quark couplings must nevertheless be taken with some care. 
In fact, it is clear that a dominance of a specific four-quark coupling does neither guarantee the 
formation of an associated condensate in the IR nor does it exclude the formation of other condensates. 
Still, it should also be mentioned that corresponding studies 
in the context of condensed-matter theory show that the appearance of a clear dominance of 
a specific four-fermion coupling is indeed a precursor for the formation of the 
associated condensate~\cite{Roscher:2019omd}.

In order to probe the effect of explicit $\UAone$-symmetry breaking on the phase boundary and 
the dominance pattern of the four-quark couplings, we implemented 
$\UAone$-violating initial conditions in form of finite values for 
the coupling associated with the so-called \textit{'t Hooft} channel. 
Even at large chemical potential, the considered strengths of explicit $\UAone$ symmetry breaking 
at the initial UV scale showed surprisingly little effect on the shape of the phase boundary. 
The same is true for the actual position of the point~$\mu_\chi$, where the dominance pattern of the four-quark couplings 
changes qualitatively.  
Compared to the $\UAone$-symmetric RG flow, however, $\UAone$-violating initial conditions 
affect the ``hierarchy'' of the four-quark couplings along the phase boundary which is 
clearly visible in, e.g., the amplification of the dominance of the scalar-pseudoscalar interaction channel at small chemical potential 
as well as of the dominance of the CSC channel at large chemical potential. 
From this, we conclude that explicit $\UAone$ symmetry breaking plays indeed an important role 
in the formation of the condensates, in particular with respect to the formation of the 
conventional CSC ground state at large chemical 
potential~\cite{Alford:1997zt,Rapp:1997zu,Berges:1998rc,Son:1998uk,Pisarski:1999tv,Pisarski:1999bf,Schafer:1999jg,Brown:1999aq,Hong:1999fh,Evans:1999at}. 
Future extensions of our present work should of course include a direct computation of 
the \textit{'t~Hooft}-coupling within our RG flow by following earlier 
works in this direction~\cite{Pawlowski:1996ch}. 

Of course, our present study can be further improved in various directions.  
Still, our analysis already provides an important insight into the dynamics underlying the QCD 
phase structure at finite temperature and quark chemical potential and consolidates the findings 
obtained from our preceding \Fierz-complete NJL-type model study~\cite{Braun:2018bik}. 
Moreover, the inclusion of gauge degrees of freedom 
combined with the \Fierz-complete set of four-quark interactions enables us to identify the relevant effective 
low-energy degrees 
of freedom and to determine, or at least constrain, from first principles 
the couplings of a suitably constructed 
truncation for the low-energy sector, in particular at large quark chemical potential. 
In the future, this may prove very valuable, e.g., to study the thermodynamics of quark matter at high density. 
Indeed, first steps into this direction have already been taken very recently~\cite{Leonhardt:2019fua}.

%%%%%%%%%%%%%%%%%%%%%%%%%%%%%%%%%%%%%%%%%%%%%%%%%%%%%%%%
{\it Acknowledgments.--~} The authors would like to thank Markus Q. Huber, J.~M.~Pawlowski, and D.~Rosenbl\"uh 
for useful discussions and J.~M.~Pawlowski also for comments on the manuscript.
This work has been done within the {\it fQCD collaboration}~\cite{fQCD}.  
J.B. acknowledges support by the DFG under Grant No. BR 4005/4-1 (Heisenberg program).
J.B. acknowledges support by HIC for FAIR within the LOEWE program of the State of Hesse. 
This work is supported by the DFG through grant SFB~1245.
%%%%%%%%%%%%%%%%%%%%%%%%%%%%%%%%%%%%%%%%%%%%%%%%%%%%%%%%

%
\bibliography{qcd}

\end{document}